\documentclass{article}
\newcommand{\tr}{\mathop{\rm Tr}\nolimits}
\newcommand{\re}{\mathop{\rm Re}\nolimits}
\newcommand{\br}{\langle}
\newcommand{\kt}{\rangle}
\newcommand{\vep}{\varepsilon}
\newcommand{\ptl}{\partial}
\newcommand{\nn}{\mu_{\rm{B}}}
\newcommand{\sign}{\mathop{\rm sgn}\nolimits}
\newcommand{\supp}{\mathop{\rm supp}\nolimits}

\begin{document}
\author{D.P.Sankovich
\thanks{Steklov Institute of Mathematics, RAS, Moscow, Russia;
E-mail: sankovch@mi.ras.ru
}}
\title{
Bogolyubov Measure in Quantum Equilibrium Statistical Mechanics
}

\date{}
\maketitle

\begin{abstract}
Application of the functional integration methods in equilibrium
statistical mechanics of quantum Bose-systems is considered.
We show that Gibbs equilibrium averages of Bose-operators can be represented
as path integrals over a special Gauss measure defined in the corresponding
space of continuous functions. This measure arises in the Bogolyubov
$T$-product approach and is non-Wiener.
We consider problems related to integration with respect to the Bogolyubov
measure in the space of continuous functions and calculate some functional
integrals with respect to this measure. Approximate formulas that are exact
for functional polynomials of a given degree and also some formulas that are
exact for integrable functionals belonging to a broader class are
constructed.
We establish the nondifferentiability of the Bogolyubov trajectories in
the corresponding function space and prove a theorem on the quadratic
variation of trajectories and study the properties implied by this theorem
for the scale transformations. We construct some examples of semigroups
related to the Bogolyubov measure. Independent increments are found for
this measure. We consider the relation between the Bogolyubov measure and
parabolic partial differential equations.
An inequality for some traces is proved, and an upper estimate is
derived for the Gibbs equilibrium mean square of the coordinate operator in
the case of a one-dimensional nonlinear oscillator with a positive symmetric
\end{abstract}

\section {Introduction}

The purpose of this article is to provide a mathematical treatment of the
Bogolyubov functional integral and to introduce some possible applications
of this integral to the equilibrium quantum statistical mechanics.

Studying integration problems for functions on an abstract set was initiated
by Fr\'echet~\cite{1a2}, who appropriately generalized the Lebesgue
method.  Somewhat later, these problems were studied by
Daniell~\cite{2a2,3a2}, who used the idea of extending linear
functionals. The Daniell theory is based on the family $H(X)$ of
elementary functions $h(x)$ on a set $X$ with an elementary integral
$I(h)$ defined for them. Under some conditions, this family can be
extended to a broader family $L$ to which the integral $I$ is extended
such that $L$ becomes a Banach space with the norm
$\|\varphi\|=I\bigl(|\varphi|\bigr)$.  This is the essence of the
construction of the Lebesgue integral in the Daniell
scheme~\cite{4a2,5a2}.

The early results by Wiener~\cite{6a2} have much in common with the theory
of the Daniell integral. He defined the integration process for
functionals and showed that the integral he considered is the Daniell
integral. We note that from 1921 on, the problem of functional integration
in all works by Wiener is related to studying the Brownian motion of
particles. The set $C=C[0,1]$ of continuous real functions $x(t)$
satisfying the condition $x(0)=0$ is defined on the interval $[0,1]$,
where $x(t)$ is the coordinate of a particle issuing from the origin at
$t=0$ and undergoing Brownian motion along the $x$ axis under the action
of random impulses. The Wiener measure has a zero mean and a correlation
function $\min(t,s)$. This measure belongs to a more general class of
measures called Gaussian measures.

Feynman~\cite{7a2} was the first to use functional integration in quantum
physics. The construction of the Feynman functional (continual) integral has
some properties in common with the Wiener integral. However, these integrals
are essentially different~\cite{8a2}.

The idea of writing physical observables as continual integrals was developed
in quantum field theory for representing the Green's function. In due course,
two such representation methods appeared almost simultaneously. One of them
was based on formal integration of equations in variational derivatives for
Green's functions~\cite{9a2}--\cite{12a2}. Bogolyubov developed a different
approach~\cite{13a2} proceeding from the representation of Green's
functions in terms of vacuum expectations of chronological products, and
the averaging operation over the boson vacuum was interpreted as a
functional integral.  In~\cite{14a2}, the Bogolyubov functional
integration method was used to study problems of gradient transformations
for electrodynamic Green's functions and to investigate the
Bloch--Nordsiek model. Bogolyubov returned to this construction in the
framework of statistical mechanics to investigate the polaron
model~\cite{15a2}. It was shown in~\cite{16a2} that the measure appearing
in the Bogolyubov approach is the Gaussian measure in the related space of
continuous functions. The Gibbs equilibrium means of chronological
products of operators are expressed in the form of functional integrals
with respect to this measure.

In Section 2, conception of the $T$-product is considered and the
Bogolyubov measure is introduced.
In Section 3, the main results of the integration theory in abstract
spaces as applied to the specific case of the Bogolyubov measure are
presented.
In Section 4, some simplest functional integrals with respect
to the Bogolyubov measure are calculated.
In Section 5, formulas of approximate integration are considered.
In Section 6, we give a brief discussion of a probabilistic approach to
the Bogolyubov process.
In Section 7, some properties of the Bogolyubov trajectories are studied
and scale transformations in the Bogolyubov space are considered.  In
Section 8, examples of semigroups related to the Bogolyubov measure are
constructed, independent increments for this measure are found and
relation between the Bogolyubov measure and parabolic partial
differential equations is considered.
In Section 9, an inequality for traces that is used in phase
transition theory is proved.

\section{Gaussian functional integrals and Gibbs equilibrium averages}
\subsection{$T$-product}

The notion of the chronological product ($T$-product) of operators appeared
in quantum mechanics in the analysis of the Schr\"odinger equation with a
time-dependent Hamiltonian~\cite{1a1}. This equation emerges in the
so-called interaction representation and is
$$
i\frac{d\Phi(t)}{dt}=\widetilde H(t)\Phi(t),
$$
where
$$
\widetilde H(t)=e^{iH_0(t-t_0)}Ve^{-iH_0(t-t_0)}
$$
and $H=H_0+V$ is the time-independent Hamiltonian of the dynamic system under
consideration. If $\Phi$ is a time-independent state vector in the Heisenberg
representation, then $\Phi(t)=S(t,t_0)\Phi$, where
$$
S(t,t_0)=e^{iH_0(t-t_0)}e^{-iH(t-t_0)},\qquad S(t_0,t_0)=I.
$$
The evolution operator $S(t,t_0)$ satisfies the equation and the initial
condition
$$
i\frac{\partial}{\partial t}S(t,t_0)=\widetilde H(t)S(t,t_0),\qquad
S(t_0,t_0)=I.
\eqno (1)
$$
In quantum mechanics, the operator $S(+\infty,-\infty)$ is called the
scattering matrix~\cite{2a1}. The evolution operator $S(t,t_0)$ is a
unitary propagator~\cite{3a1}, i.e., it satisfies the conditions that

\begin{itemize}
\item[a)] $S(t,t_1)S(t_1,t_0)=S(t,t_0)$,
\item[b)] $S(t,t)=I$, and
\item[c)] $S(t,t_0)$ is strongly continuous in all the variables $t$ and
$t_0$.
\end{itemize}

The equation with initial condition~(1) is formally equivalent to the
integral equation
$$
S(t,t_0)=I-i\int_{t_0}^t \widetilde H(\tau)S(\tau,t_0)\,d\tau.
$$
Using consecutive substitutions, we can establish the Dyson expansion
$$
S(t,t_0)=\sum_{n=0}^\infty S_n(t,t_0),
\eqno(2)
$$
where
$$
S_n(t,t_0)=(-i)^n\int_{t_0}^t\,dt_1
\int_{t_0}^{t_1}dt_2\dots\int_{t_0}^{t_{n-1}}dt_n\,
\widetilde H(t_1)\widetilde H(t_2)\dots\widetilde H(t_n).
\eqno(3)
$$
It is convenient to write this as
$$
S_n(t,t_0)=\frac{(-i)^n}{n!} \int_{t_0}^t dt_1\,
\int_{t_0}^tdt_2\dots\int_{t_0}^tdt_n\,T\bigl[\widetilde
H(t_1)\widetilde H(t_2)\dots\widetilde H(t_n)\bigr],
$$
where we introduce the $T$-product
$$
T[V(t_1)V(t_2)\dots V(t_n)]=\sum\pm\theta(t_{i_1}>t_{i_2}>\dots>
t_{i_n})V(t_{i_1})V(t_{i_2})\dots V(t_{i_n}),
\eqno(4)
$$
with
$$
\theta(t_1>t_2>\dots>t_n)=
\cases{
1\quad &if $t_1\geq t_2\geq\dots\geq t_n$,
\cr
0\quad & otherwise.
\cr
}
$$

The sum in Eqs.~(4) is taken over all possible permutations of the
indices $1,2,\dots,n$. The minus sign corresponds to the Fermi case and is
determined by the number of Fermi transpositions that are necessary for the
derivation of the corresponding term.

Using Eqs.~(4), we can write expansion~(2) in the symbolic form
$$
S(t,t_0)=T\exp\biggl[\,-i\int_{t_0}^t\widetilde H(\tau)\,d\tau\biggr].
$$
It follows from the definition of the $T$-product that operators commute
under the $T$-product sign.

The general conditions for the existence of the solution of the evolution
equation
$$
\frac{d\varphi(t)}{dt}=A(t)\varphi(t)
$$
with an unbounded operator $A(t)$ were first found in~\cite{4a1}.

Using the $T$-product, we can obtain an important formula of equilibrium
statistical mechanics~\cite{15a2}. We consider the operator equation
$$
\frac{dU(s)}{ds}=-[H_0+H_1(s)]U(s),\qquad U(0)=I,
\eqno(5)
$$
which is solved by
$$
U(\beta)=T\exp\left\{-\int_0^\beta\bigl[H_0+H_1(\sigma)\bigr]\,
d\sigma\right\}.
\eqno(6)
$$
We assume that $U(s)=e^{-sH_0}C(s)$ in~(5). Then the equation and the
initial condition satisfied by $C(s)$ are
$$
\frac{dC(s)}{ds}=-e^{sH_0}H_1(s)e^{-sH_0}C(s),\qquad C(0)=I
$$
and are solved by
$$
C(s)=T\exp\biggl[\,-\int_0^s d\sigma\,
e^{\sigma H_0}H_1(\sigma)e^{-\sigma H_0}\biggr].
$$
Therefore,
$$
U(\beta)=e^{-\beta H_0}T\exp\biggl[\,-\int_0^\beta ds\,
e^{sH_0}H_1(s)e^{-sH_0}\biggr].
\eqno(7)
$$
For the special case where the operator $H_1(s)=H_1$ is independent of $s$,
we compare solutions~(6) and~(7) and thus obtain the Bogolyubov
formula
$$
e^{-\beta(H_0+H_1)}=e^{-\beta H_0}T\exp\biggl[\,-\int_0^\beta ds\,
e^{sH_0}H_1e^{-sH_0}\biggr].
$$
This formula is necessary for representing the partition function as a path
integral.

\subsection{ Gibbs equilibrium averages}

If $\widehat A$ is a linear span of Bose operators and $\widehat\Gamma$ is a
positive-definite quadratic Hamiltonian, we have the formula~\cite{15a2}
$$
2\ln\langle e^{\widehat A}\rangle=\langle{\widehat A}^2\rangle,
\eqno(8)
$$
where
$$
\langle\cdot\rangle=\frac{{\tr}\bigl[\,\cdot\,e^{-\beta\widehat\Gamma}\bigr]}
{\tr\,e^{-\beta\widehat\Gamma}}
$$
denotes the Gibbs average with the Hamiltonian~$\widehat\Gamma$.

We consider the average
$$
\Biggl\langle T\exp\biggl[\,i\sum_{k=1}^{N+1}\nu_k
\widehat Q(s_k)\biggr]\Biggr\rangle,
\eqno(9)
$$
where $\nu_k$ are real numbers and
$$
0=s_1<s_2<\dots<s_k<\dots<s_N<s_{N+1}=\beta.
\eqno(10)
$$
The operators $\widehat Q(s)$ and $\widehat\Gamma$ are given by
$$
\widehat Q(s)=e^{s\widehat\Gamma}\hat qe^{-s\widehat\Gamma}, \qquad
\widehat\Gamma=\frac{\hat p^2}{2m}+\frac{m\omega^2}{2}\hat q^2,
$$
which means that we consider the one-dimensional harmonic oscillator. Taking
Eq.~(8) into account, we can write
$$
\Biggl\langle T\exp\biggl[\,i\sum_{k=1}^{N+1}\nu_k
\widehat Q(s_k)\biggr]\Biggr\rangle
=\exp\left\{-\frac12\sum_{n=1}^{N+1}\sum_{m=1}^{N+1}
\nu_n\nu_m\Bigl\langle T\bigl[\widehat Q(s_n)\widehat Q(s_m)\bigr]
\Bigr\rangle\right\}.
$$
We evaluate the average in the right-hand side of the last relation using
$T$-product definition~(4), which leads us to
$$
\Bigl\langle T\bigl[\widehat Q(s_n)\widehat Q(s_m)\bigr]\Bigr\rangle=
\bigl(2m\omega(1-e^{-\beta\omega})\bigr)^{-1}\bigl(e^{-\omega|s_n-s_m|}+
e^{-\beta\omega+\omega|s_n-s_m|}\bigr).
$$
Thus, average~(9) can be represented as
$$
\Biggl\langle T\exp\biggl[\,i\sum_{k=1}^{N+1}\nu_k
\widehat Q(s_k)\biggr]\Biggr\rangle=
$$
$$
=\exp\biggl[\,-\frac12\sum_{n,m=1}^{N+1}\nu_n\nu_m
\bigl(2m\omega(1-e^{-\beta\omega})\bigr)^{-1}
\bigl(e^{-\omega|s_n-s_m|}+
e^{-\beta\omega+\omega|s_n-s_m|}\bigr)\biggr].\\
$$

We now write the last formula in a more convenient form for the future
analysis. We consider the expression
$$
K(s_n,s_m)=e^{-\omega|s_n-s_m|}+e^{-\beta\omega+\omega|s_n-s_m|},
\quad 0<s_n,s_m<\beta,
$$
as a function of $s_n$. This function, which we represent by~$y(s_n)$,
satisfies the differential equation
$$
\frac{d^2y(s_n)}{ds_n{}^2}-{\omega}^2y(s_n)=
-2\omega\left(1-e^{-\beta\omega}\right)\delta(s_n-s_m)
\eqno(11)
$$
and the boundary conditions
$$
y(0)=y(\beta),\qquad y^\prime (0)=y^\prime(\beta).
$$
We seek the solution of Eq.~(11) in the form
$$
y(s)=\sum_{n=-\infty}^\infty c_ne^{2\pi ins/\beta}.
$$
It follows that
$$
K(s_j,s_k)=2\omega\frac{1-e^{-\beta\omega}}{\beta}
\sum_{n=-\infty}^\infty\frac{e^{2\pi in(s_j-s_k)/\beta}}
{\omega^2+(2\pi n/\beta)^2}.
$$
For average~(9), we thus have the representation
$$
\Biggl\langle T\exp\biggl[\,i\sum_{k=1}^{N+1}\nu_k\widehat
Q(s_k)\biggr]\Biggr\rangle=e^{-\Omega(\{\nu_k\})}
\eqno(12)
$$
with the quadratic form in $\nu_k$ given by
$$
\Omega(\{\nu_k\})\equiv\frac1{2m\beta}\sum_{n=-\infty}^\infty
\frac{\bigl|\sum_{k=1}^{N+1}\nu_ke^{2\pi ins_k/\beta}\bigr|^2}
{\omega^2+(2\pi n/\beta)^2}.
$$
Obviously, $\Omega\geq 0$. In addition, $\Omega=0$ if and only if
$\nu_1+\nu_{N+1}=0$ and $\nu_2=0$, \dots, $\nu_N=0$.

Introducing new variables $\eta_1=\nu_1+\nu_{N+1}$, \ $\eta_2= \nu_2$,
\dots, $\eta_N=\nu_N$, we can rewrite Eq.(12) as
$$
\Biggl\langle T\exp\biggl[\,i\sum_{k=1}^{N+1}\nu_k\widehat Q(s_k)
\biggr]\Biggr\rangle=\exp\biggl(-\frac{1}{2}\sum_{j=1}^N\sum_{k=1}^N
A_{jk}\eta_j\eta_k\biggr),
\eqno(13)
$$
where
$$
\sum_{j,k=1}^N A_{jk}\eta_j\eta_k=
\frac1{m\beta}\sum_{n=-\infty}^\infty
\frac{\bigl|\sum_{k=1}^N\eta_ke^{2\pi ins_k/\beta}\bigr|^2}
{\omega^2+(2\pi n/\beta)^2}
\eqno(14)
$$
and the covariance matrix entries are
$$
A_{jk}=\frac1{2m\omega
\sinh(\beta\omega/2)}\cosh\biggl(\frac{\beta\omega}2-
\frac{\beta\omega}N |j-k|\biggr).
$$
In deriving the last formula, a partition of form~(10) was defined by
the simple relation~$s_j=\beta N^{-1}(j-1)$.

We now apply Eqs.~(13) and~(14) to find the relation between
the Gibbs equilibrium averages of Bose operators and the path integral.

\subsection {Gaussian path integrals}

We consider the expression
$$
\int\Biggl\langle T\exp\biggl[\,i\sum_{k=1}^{N+1}\nu_k
\widehat Q(s_k)\biggr]\Biggr\rangle
\exp\biggl\{-i\sum_{k=1}^{N+1}\nu_kq_k\biggr\}\,
d\nu_1\dots d\nu_N\,d\nu_{N+1},
$$
where $q_k$ are real numbers and the integration with respect to each
variable $\nu_i$ goes over the entire real axis. Taking Eq.~(13) and
the known values of Gaussian integrals into account, we obtain
$$
\frac{1}{(2\pi)^{N+1}}\int\Biggl\langle
T\exp\biggl[\,i\sum_{k=1}^{N+1}\nu_k
\widehat Q(s_k)\biggr]\Biggr\rangle
\exp\biggl\{-i\sum_{k=1}^{N+1}\nu_kq_k\biggr\}\,
d\nu_1\dots d\nu_N\,d\nu_{N+1}=
$$
$$
=\rho(q_1,q_2,\dots,q_{N+1}),
\eqno(15)
$$
where
$$
\rho(q_1,q_2,\dots,q_{N+1})=
\frac1{\sqrt{(2\pi)^N}}\frac{\delta(q_1-q_{N+1})}{\sqrt{\det A}}
\exp\biggl[\,-\frac12\sum_{j,k=1}^N(A^{-1})_{jk}q_jq_k\biggr],
\eqno(16)
$$
$\delta(q)$ is the Dirac delta function, and $A^{-1}$ is the
inverse covariance matrix with the entries
$$
\bigl(A^{-1}\bigr)_{ij}=\frac{m\omega}{\sinh(\beta\omega/N)}
\biggl(2\cosh\frac{\beta\omega}N\delta_{i,j}
-\delta_{i,j+1}-\delta_{i,j-1}\biggr).
$$
The determinant of the inverse covariance matrix is
$$
\det\bigl(A^{-1}\bigr)=\frac12
(m\omega)^N\biggl[\biggl(\cosh\frac{\beta\omega}N+1\biggr)^{N+1}
-\biggl(\cosh\frac{\beta\omega}N-1\biggr)^{N+1}\biggr].
$$
It follows from~(16) that
$$
\rho\geq0,\qquad\int\rho\,dq_1\dots dq_{N+1}=1.
\eqno(17)
$$

Using relation~(15), we can evaluate the averages of the form
$$
\bigl\langle T\bigl[f(\widehat Q(s_1),\dots,
\widehat Q(s_{N+1}))\bigr]\bigr\rangle.
$$
Indeed, we recall the complex Fourier formula
$$
f(Q_1,\dots,Q_{N+1})=
\frac1{(2\pi)^{N+1}}
\int f(q_1,\dots,q_{N+1})\times
$$
$$
\times\exp\biggl\{i\sum_{j=1}^N\nu_j(Q_j-q_j)\biggr\}\,
dq_1\dots dq_{N+1}\,d\nu_1\dots d\nu_{N+1}.
$$
Because the operators $\widehat Q(s_j)$ commute under the $T$-product sign,
we have
$$
\bigl\langle T\bigl[f\bigl(\widehat Q(s_1),\dots,
\widehat Q(s_{N+1})\bigr)\bigr]\bigr\rangle=
\int f(q_1,\dots,q_{N+1})\rho(q_1,\dots,q_{N+1})\,dq_1\dots dq_{N+1}.
\eqno(18)
$$
Now using properties~(17), we see that
$$
0\leq\bigl\langle T\bigl[f\bigl(\widehat Q(s_1),\dots,
\widehat Q(s_{N+1})\bigr)\bigr]\bigr\rangle\leq M,\;\;
\mbox{if }0\leq f\bigl(\widehat Q(s_1),\dots,
\widehat Q (s_{N+1})\bigr)\leq M.
\eqno(19)
$$

We now consider the functionals $F(q)$ of real functions (``trajectories")
$q(s)$ defined on the segment $0\leq s\leq\beta$. We construct the integral
$$
I\equiv\int F(q)\,d\mu
\eqno(20)
$$
over the corresponding measure.

We first consider the subset of ``special functionals''~\cite{15a2} that are
continuous functions of a finite number $N$ of variables,
$$
F^{(N)}(q)\equiv\Phi(q_1,q_2,\dots,q_N),
$$
where $q_j=q(s_j)$. By definition, we then have
$$
I^{(N)}=\int\Phi(q_1,q_2,\dots,q_N)\rho(q_1,q_2,\dots,q_N)\,
dq_1\,dq_2\dots dq_N.
\eqno(21)
$$
It follows then from~(18) and~(19) that
$$
\bigl\langle T\bigl[F^{(N)}(\widehat Q)\bigr]\bigr\rangle=
\int F^{(N)}(q)\,d\mu
$$
and $\bigl\langle T\bigl[F^{(N)}(\widehat Q)\bigr]\bigr\rangle\geq0$ if
$F^{(N)}(q)\geq 0$ for arbitrary real numbers $q_1,q_2,\dots,q_N$. We now
consider the sequence of functions $\{q_N(s)\}$, $N=1,2,\dots$, defined as
$$
q_N(s)=q(s_j)\qquad\mbox{for}\quad s_j\leq s< s_{j+1},\quad
j=1,2,\dots,N,\qquad q_N(\beta)=q(\beta).
\eqno(22)
$$
The set of points $\{s_j\}$ is the partition~(10) of the segment
$[0,\beta]$. We assume that $|s_{j+1}-s_j|\leq\Delta s$ for $j=1,2,\dots,N$
and also that $\Delta s\to 0$ as $N\to\infty$. Then the sequence of
step-functions~(22) uniformly tends to the function $q(s)$. Path
integral~(20) can be defined as the $N\to\infty$ limit of integrals
~(21), which are defined on the subset of ``special functionals,"
because the functionals $F(q_N(s))$ belong to this subset; therefore,
$$
I=\lim_{N\to\infty}I^{(N)}.
$$

We consider the space $C^\circ[0,\beta]$ of continuous functions $q(s)$
defined on the segment $[0,\beta]$ that satisfy the condition $q(0)=
q(\beta)$. This is a metric space with respect to the uniform metric
$$
\rho(q,p)=\sup_{s\in[0,\beta]}|q(s)-p(s)|.
$$
The square order-$N$ matrix $A=(A_{jk})$ is positive and symmetrical, i.e.,
the mapping $(j,k)\to A_{jk}$ is a positive-type kernel on the set
$\{1,2,\dots,N\}$. Therefore~\cite{6a1}, we can speak of the Gaussian
measure $\gamma_A$ on the space $R^N$ with the covariance~$A$. By the
Stone--Weierstrass theorem \cite{5a2}, the corresponding set of ``special
functionals" is dense in the set of all continuous functions defined on the
space $C^\circ[0,\beta]$. In $C^\circ[0,\beta]$, we can introduce a
$\sigma$-algebra generated by quasi intervals (cylindrical sets). This
$\sigma$-algebra is the same as the $\sigma$-algebra generated by the sets
that are open in the metric~$\rho$. Extending the Gaussian measure from the
quasi intervals to their Borel closure, we obtain a Gaussian measure in the
space $C^\circ[0,\beta]$ \cite{8a1}.

\section {The Bogolyubov measure in the space of continuous functions}

So we see that the Gaussian measure $\nn$ with zero
average and the correlation function
$$
B(t,s)=\frac1{2m\omega\sinh(\beta\omega/2)}
\cosh\left(\omega|t-s|-\frac{\beta\omega}2\right)
\eqno (23)
$$
is defined in the space $X=C^{\circ}[0,\beta]$ of continuous functions on the
interval $[0,\beta]$ with the uniform metric $\rho=\max_{t\in[0,\beta]}
\bigl|x(t)-y(t)\bigr|$ that satisfy the condition $x(0)=x(\beta)$. Measurable
functionals $F(x)$ are considered on the space with measure
$\{X,G,\nn\}$, where $G$ is an isolated $\sigma$ algebra of subsets in
this space. In this case, the formula
$$
\bigl\langle T\bigl[F\bigl(\widehat
Q(t)\bigr)\bigr]\bigr\rangle_{\widehat{\Gamma}}=
\int_XF\bigl(x(t)\bigr)\,d\nn(x)
\eqno (24)
$$
holds for the Gibbs equilibrium mean of the $T$-product taken with respect to
the Hamiltonian $\widehat\Gamma$ of the harmonic oscillator; the integral
is understood as the Daniell integral over the space $X$,
$$
\widehat\Gamma=\frac{\hat p^2}{2m}+\frac{m\omega^2}2\hat q^2,\qquad
\widehat Q(t)=e^{t\widehat{\Gamma}}\hat qe^{-t\widehat{\Gamma}},\qquad
\langle\,\cdot\,\rangle_{\widehat{\Gamma}}=
\frac{\tr(\,\cdot\,e^{-\beta\widehat{\Gamma}})}{\tr
e^{-\beta\widehat{\Gamma}}},
$$
where $\hat q$ and $\hat p$ are the
respective coordinate and momentum operators of a particle with mass $m$
that satisfy the commutation relation $[\hat q,\hat p]=i$ \ ($\hbar=1$ is
assumed), $\beta$ is the reciprocal of the temperature, and $\omega$ is
the eigenfrequency of the oscillator ($\beta>0$, $\omega>0$.) The mean in
formula~(24) exists and is finite for an integrable functional
$F(x)$. The measure $\nn$ thus defined is called the {\sl Bogolyubov
measure}.

The kernel $B(t,s)$ of the correlation operator $B$ is symmetric and
Hermitian. It belongs to the space $L^2$ of square summable functions of two
variables with respect to the Lebesgue measure in the domain
$0\leq t\leq\beta$, $0\leq s\leq\beta$. By the Schmidt theorem~\cite{5a2},
every square summable function $A(t,s)$ that is symmetric with respect to its
arguments can be expanded as a series
$$
A(t,s)=\sum_n\lambda_n\Phi_n(t)\overline{\Phi_n(s)}
\eqno(25)
$$
in the sense of the convergence in the mean, where $\bigl\{\Phi_n(t)\bigr\}$
is an orthonormalized sequence of eigenfunctions and $\{\lambda_n\}$ is the
sequence of the corresponding eigenvalues of the operator $A$ generated by
the kernel $A(t,s)$. According to~\cite{16a2}, correlation
function~(23) has an expansion of form~(25), where
$$
\Phi_n(t)=\frac1{\sqrt{\beta}}e^{2\pi int/\beta},\qquad
\lambda_n=\frac1m\frac1{\omega^2+(2\pi n\beta^{-1})^2},
$$
and $n$ ranges the set of all integers from $-\infty$ to $\infty$. By the
Mercer theorem~\cite{5a2}, series~(25) for the kernel $B(t,s)$ is
uniformly convergent because the operator $B$ generated by $B(t,s)$ is
positive. We also note that this operator is completely continuous.
Series~(25) for the correlation function $B(t,s)$ can be written in the
space of real functions in the form
$$
B(t,s)=\sum_{n=-\infty}^{+\infty}\lambda_n\varphi_n(t)\varphi_n(s),
$$
where
$$
\varphi_n(t)=\cases{
\sqrt{\frac2\beta}\cos\frac{2\pi nt}\beta, &$n>0$; \cr
\sqrt{\frac2\beta}\sin\frac{2\pi nt}\beta, &$n<0$; \cr
\sqrt{\frac1\beta}, &$n=0$. \cr}
$$

The conjugate space of $X$, $X'=V_0[0,\beta]$, is the space of functions of
bounded variation on $[0,\beta]$ that satisfy the conditions
$$
g(0)=0,\qquad
g(t)=\frac12\bigl[g(t+0)+g(t-0)\bigr]\quad\mbox{for }t\in(0,\beta).
$$
By the Riesz representation theorem~\cite{5a2}, the linear functionals in
$X$ have the form
$$
\langle\varphi,x\rangle=\int_0^{\beta} x(t)\,d\varphi(t),
$$
where the integral is understood as the Stieltjes integral, $x(t)\in X$, and
$\varphi(t)\in V_0[0,\beta]$. The correlation functional in the space
$X'$ can be written as
$$
K(\varphi,\psi)=\int_0^{\beta}\int_0^{\beta}B(t,s)\,
d\varphi(t)\,d\varphi(s),
$$
where the correlation function for the measure has the form
$$
B(t,s)=\int_Xx(t)x(s)\,d\mu(x).
$$
By the Kuelbs theorem~\cite{21a2}, the Hilbert space $H$ generated by the
measure $\mu$ is the linear span of the eigenfunctions
$\bigl\{\varphi_n(t)\bigr\}$ of the kernel $B(t,s)$. This linear span is
closed with respect to the norm corresponding to the inner product
$$
(x,y)_H=\sum_{n=-\infty}^{+\infty}\frac1{\lambda_n}
\biggl(\,\int_0^{\beta}x(t)\varphi_n(t)\,dt\biggr)
\biggl(\,\int_0^{\beta} y(t)\varphi_n(t)\,dt\biggr).
$$
The functions $\bigl\{e_n(t)=\sqrt{\lambda_n}\,\varphi_n(t)\bigr\}_{n=
-\infty}^{+\infty}$ form a basis in the space $H$, and the expansion
$$
x(t)=\sum_{n=-\infty}^{+\infty}
\biggl(\,\int_0^{\beta}x(t)\varphi_n(t)\,dt\biggr)\varphi_n(t)
$$
holds for almost all $x\in X$. The general form of a linear measurable
functional on $X$ is given by the expression
$$
(a,x)=\sum_{n=-\infty}^{+\infty}\frac1{\lambda_n}
\biggl(\,\int_0^{\beta}x(t)\varphi_n(t)\,dt\biggr)
\biggl(\,\int_0^{\beta}a(t)\varphi_n(t)\,dt\biggr),
$$
where $a\in H$ and $x\in X$. The functions
$$
e_n(t)=\int_0^{\beta}B_{1/2}(t,u)\alpha_n(u)\,du
$$
also form a complete orthonormalized system in $H$, where $B_{1/2}(t,u)$ is
the kernel of the operator $B^{1/2}$ and $\alpha_n(t)$ is an arbitrary
complete orthonormal system in the space $L_2[0,\beta]$.

We note that the closure $\overline H$ of the Hilbert space $H$ is the
support of the measure $\mu$ and is dense almost everywhere in
$X$~\cite{22a2}.  The triple $(X,H,\mu)$ is called an {\sl abstract Wiener
space}, and the measure $\mu$ is called an {\sl abstract Wiener
measure}~\cite{23a2}.

We also note that in the case of the Bogolyubov measure, $G(t,s)=-mB(t,s)$ is
the Green's function of the boundary value problem
$$
\left\{
\begin{array}{l}
y''-\omega^2 y=0,\\
y(0)=y(\beta),\\
y'(0)=y'(\beta)\\
\end{array}
\right.
$$
on the interval $[0,\beta]$.

\section {Functional integral with respect
to the Bogolyubov measure}

Let $a_1,a_2,\dots,a_n$ be linearly independent elements in a separable
Hilbert space $H$ whose closure is the support of a measure $\mu$ and which
is dense almost everywhere in $X$. Then
$$
\int_XF\bigl[(a_1,x),(a_2,x),\dots,(a_n,x)\bigr]\,d\mu(x)=
$$
$$
=(2\pi)^{-n/2}\frac1{\sqrt{\det A}}\int_{R^n}e^{-(A^{-1}u,u)/2}F(u)\,du
\eqno(26)
$$
if one of the integrals in~(26) exists, where $A$ is the matrix of the
elements $a_{ij}=(a_i,a_j)_H$, $i,j=1,2,\dots,n$, $u=(u_1,u_2,\dots,u_n)$,
and $du=du_1du_2\cdots du_n$. If orthonormalized vectors in $H$ are taken as
the elements $a_j$, then~(26) becomes
$$
\int_XF\bigl[(a_1,x),(a_2,x),\dots,(a_n,x)\bigr]\,d\mu(x)=
(2\pi)^{-n/2}\int_{R^n}e^{-(u,u)/2}F(u)\,du.
$$
The form of formula~(26) is particularly simple if the functional
$F(x)$ depends only on the values of the function $x(t)$ at finitely many
points. For example, if
$$
F\bigl(x(t)\bigr)=x(t_1)x(t_2)\cdots x(t_n),
$$
then the Wick theorem holds, by which
$$
\int_X x(t_1)x(t_2)\cdots x(t_n)\,d\mu(x)=
\sum B(t_{i_1},t_{i_2})B(t_{i_3},t_{i_4})\cdots B(t_{i_{2k-1}},t_{i_{2k}}),
$$
where $n=2k$ and the summation extends over all $(2k)!/(2^kk!)$
decompositions of the numbers $1,2,\dots,2k$ into $k$ different unordered
pairs,
$$
(i_1,i_2),(i_3,i_4),\dots,(i_{2k-1},i_{2k}).
$$
This integral vanishes for $n=2k+1$. In particular, for the case of the
Bogolyubov measure, we have
$$
\langle\hat q^2\rangle_{\widehat{\Gamma}}=
\int_Xx^2(t)\,d\nn(x)=B(t,t)=
\frac1{2m\omega}\coth\frac{\beta\omega}2,
$$
$$
\langle e^{a\hat q^2}\rangle_{\widehat{\Gamma}}=
\sum_{n=0}^{\infty}\frac{(2n)!}{2^n(n!)^2}
\bigl(a\langle\hat q^2\rangle_{\widehat{\Gamma}}\bigr)^n=
\frac1{\sqrt{1-a\coth(\beta\omega/2)/(m\omega)}\,},
$$
where it is necessary to assume that $-m\omega\tanh(\beta\omega/2)\leq
a<m\omega\tanh(\beta\omega/2)$ in the second formula.

We consider the quadratic functional
$$
A(x,x)=\sum_{k,j=1}^{\infty}a_{kj}(e_k,x)(e_j,x)
$$
on $X$, where $a_{kj}=(Ae_k,e_j)_H$, $A$ is a self-adjoint kernel operator
from $H$ into $H$, and $\{e_k\}_{k=1}^{\infty}$ is a basis in $H$. Using
formula~(26), we can then calculate the integrals
$$
\int_XA(x,x)\,d\mu(x)=\tr A,\quad
\int_XA^2(x,x)\,d\mu(x)=(\tr A)^2+2\sum_{k=1}^{\infty}\lambda_k^2,
$$
where $\lambda_k$ are the eigenvalues of the operator $A$. The relation
$$
\int_Xe^{\lambda A(x,x)/2}\,d\mu(x)=\frac1{\sqrt{D_A(\lambda)}}
\eqno(27)
$$
also holds, where $D_A(\lambda)$ is the characteristic determinant of $A$ at
the point $\lambda$,
$$
\re\lambda<\frac1{\lambda_1},\qquad\lambda_1>\lambda_2>\ldots,
$$
and
$$
\sqrt{D_A(\lambda)}\,=\sqrt{\bigl|D_A(\lambda)\bigr|}\,
\exp\left[-\frac i2\arg D_A(\lambda)\right].
$$

We take
$$
A(x,x)=\int_0^{\beta}x^2(t)\,dt=
\sum_{k=-\infty}^{\infty}\lambda_k(e_k,x)^2
$$
as the quadratic functional in~(27), where $e_k=\sqrt{\lambda_k}\,
\varphi_k$, $\lambda_k$ are the eigenvalues of the kernel $B(t,s)$, and
$\varphi_k$ are the corresponding eigenfunctions. We use the
formula~\cite{5a2}
$$
-\frac d{d\lambda}\ln D_B(\lambda)= \int_0^{\beta}
B(t,t)\,dt+\lambda\int_0^{\beta} B^{(2)}(t,t)\,dt+\cdots+
\lambda^k\int_0^{\beta} B^{(k)}(t,t)\,dt+\ldots,
$$
where $B^{(k)}$ are the corresponding iterated kernels, which have the form
$$
B^{(k)}(t,t)=\frac1{\beta m^k}\sum_{n=-\infty}^{\infty}
\frac1{\bigl[\omega^2+(2\pi n\beta^{-1})^2\bigr]^k}
$$
in the case of the Bogolyubov measure. This results in
$$
-\frac d{d\lambda}\ln D_B(\lambda)=
\frac1\lambda\sum_{n=-\infty}^{\infty}\sum_{k=1}^{\infty}
\left(\frac\lambda m\right)^k
\frac1{\bigl[\omega^2+(2\pi n\beta^{-1})^2\bigr]^k}=
$$
$$
=\frac\beta{2m}\frac1{\sqrt{\omega^2-\lambda/m}}
\coth\left(\frac{\beta}{2}\sqrt{\omega^2-\frac\lambda m}\,\right),
$$
where $\lambda<m\omega^2$. Integrating this equation, we obtain
$$
D_B(\lambda)=
\frac{\sinh^2\left(\beta\sqrt{\omega^2-\lambda/m}\,/2\right)}
{\sinh^2(\beta\omega/2)},
$$
whence follows the formula
$$
\int_X\exp\biggl(\frac{\lambda}{2}\int_0^{\beta}x^2(t)\,dt\biggr)\,
d\nn(x)=\frac{\sinh(\beta\omega/2)}
{\sinh\left(\beta\sqrt{\omega^2-\lambda/m}\,/2\right)},\quad
\lambda<m\omega^2.
\eqno(28)
$$
We note that the moments
$$
m_k=\int_XA(x,x)^k\,d\nn(x)=
\int_X\biggl(\,\int_0^{\beta} x^2(t)\,dt\biggr)^k\,d\nn(x)
$$
can be determined using formula~(28) and the relation
$$
m_{k+1}=\left.2^{k+1}\frac{d^{k+1}}{d\lambda^{k+1}}
\frac1{\sqrt{D_B(\lambda)}}\right|_{\lambda=0}.
$$

Taking the relation
$$
D_B(\lambda)=\prod_{n=-\infty}^{\infty}(1-\lambda\lambda_n)
$$
into account, we derive the following formula for the infinite product from
the above value of the Fredholm determinant of the kernel $B(t,s)$:
$$
\prod_{n=1}^{\infty}\left(1+\frac a{n^2+b^2}\right)=\frac1{\sqrt{1+a/b^2}}
\frac{\sinh\left(\pi b\sqrt{1+a/b^2}\,\right)}{\sinh(\pi b)},\quad a>-b^2.
$$

\section {Approximate calculation of functional integrals}

We consider approximate formulas that are exact for functional polynomials of
a given degree. Let $X$ be the space $C[a,b]$ of continuous functions $x(t)$
on $[a,b]$. We assume that a Gaussian measure $\mu$ with a zero
average and a
correlation function $B(t,s)$ is defined in $X$. An arbitrary continuous
functional polynomial $P_n(x)$ of degree $n$ on $C$ has the form
$$
P_n(x)=p_0+\sum_{j=1}^n\int_a^b\dots\int_a^bx(t_1)\cdots x(t_j)\,
d_{t_1\cdots t_j}^{(j)}g(t_1,\dots,t_j),
$$
where $p_0=\mbox{const}$ and the other terms are multiple Stieltjes
integrals.

{\bf Theorem 1 {\rm\cite{19a2}}.}
{\sl
Let $\nu$ be a symmetric probability measure on the Borel sets in $R$, and
let $\rho(u,t)$ be a function on $R\times[a,b]$ such that
\begin{itemize}
\item[1)] $\rho(u,t)=-\rho(-u,t)$,
\smallskip
\item[2)] $\prod_{j=1}^m\rho(u,t_j)\in L(R,\nu)$ for
$1\leq m\leq 2n+1$,
\smallskip
\item[3)]
$$
\int_{-\infty}^{\infty}\rho(u,t)\rho(u,s)\,d\nu(u)=B(t,s).
\eqno(29)
$$
\end{itemize}
Then the formula
$$
\int_C F(x)\,d\mu(x)\approx
\int_{R^n}F\bigl(\theta_n(u,\cdot)\bigr)\,d\nu_n(u),
\eqno(30)
$$
where
$$
\theta_n(u,t)=\sum_{j=1}^nc_j\rho(u_j,t),
$$
$c_j^2$ are the roots of the polynomial
$$
Q_n(z)=\sum_{k=0}^n\frac{z^{n-k}}{k!},
$$
and $\nu_n$ is a measure in $R^n$ that is an $n$-fold Cartesian product of
the measures $\nu$, is exact for functional polynomials of degree $2n$$+$$1$.
}

{\bf Theorem 2 {\rm\cite{19a2}}.
}
{\sl
Let the assumptions in Theorem~{\rm1} hold. Then the formula
$$
\int_CF(x)\,d\mu(x)\approx\frac{(-1)^n(A-n)^n}{n!}F(0)+
$$
$$
+\sum_{k=1}^n(-1)^{n-k}\frac{(A-n+k)^n}{k!\,(n-k)!}
\int_{R^k}F\bigl(\theta_k^{(n)}(u,\cdot)\bigr)\,d\nu_k(u),
\eqno(31)
$$
where
$$
\theta_k^{(n)}(u,t)=\frac1{\sqrt{A-n+k}}\sum_{j=1}^k\rho(u_j,t),\qquad
R^k=\underbrace{R\times R\times\dots\times R}_k,
$$
$$
d\nu_k(u)=d\nu(u_1)\cdots d\nu(u_k),\quad k=1,2,\dots,n,
$$
and $A$ is an arbitrary constant, is exact for all functional polynomials of
degree $2n$$+$$1$.
}

If $A=n$, then formula~(31) becomes~\cite{24a2}
$$
\int_CF(x)\,d\mu(x)\approx I_n(F),
$$
where
$$
I_n(F)=\sum_{k=1}^n(-1)^{n-k}\frac{k^n}{k!\,(n-k)!}
\int_{R^k}F\bigl(\theta_k(u,\cdot)\bigr)\,d\nu_k(u)
$$
and
$$
\theta_k(u,t)=\frac1{\sqrt{k!}}\sum_{j=1}^k\rho(u_j,t).
$$
It is easy to verify that the recursive relation
$$
I_n(F)=\frac{n^n}{n!}\int_{R^n}F\bigl(\theta_n(u,\cdot)\bigr)\,d\nu_n(u)-
\sum_{k=1}^{n-1}\frac{kn^{n-1-k}}{(n-k)!}I_k(F)
$$
holds for $I_n(F)$.

Deriving formulas~(30) and~(31) for approximately calculating
integrals with respect to the Gaussian measure relates to finding a
function $\rho(u,t)$ possessing properties~{1--3} in Theorem~1. The most
difficult task here is solving Eq.~(29).

We first seek the solution of this equation for the case of a purely discrete
measure $\nu$ on the line. We recall that a measure entirely concentrated on
a finite or countable set of points on the line is said to be {\sl discrete}.

Let a finite or countable set of points $\{x_n\}_{n=-\infty}^{\infty}$ be
given on an interval $[a,b]$, and let a positive number $h_n$ satisfying the
condition
$$
\sum_nh_n<\infty
$$
be associated with each $x_n$. We define a function $f$ on $[a,b]$ by setting
$$
f(x)=\sum_{x_n<x}h_n.
$$
The function $f(x)$ does not decrease and is left-continuous. If $x$
coincides with one of the points $x_n$, with $x=x_{n_0}$ for example, then
$$
f(x_{n_0}+0)-f(x_{n_0}-0)=h_{n_0}.
$$
If $x$ does not coincide with any of the points $x_n$, then $f(x)$ is
continuous at $x$. The function $f(x)$ is called a {\sl jump function}.

We define a measure $\nu$ on $R$ in the form
$$
\nu\bigl\{(-\infty,x)\bigr\}=f(x)
$$
and assume that
$$
\sum_nh_n=1,\qquad h_n=h_{-n},\quad n=0,\pm1,\pm2,\dots\,.
$$

If we set $x_n=n$, then
$$
\int_R\rho(u,s)\rho(u,t)\,d\nu(u)=
\sum_{n=-\infty}^{\infty}h_n\rho(n,s)\rho(n,t).
$$
Expanding the correlation function in a series with respect to the complete
system of orthonormalized eigenfunctions,
$$
B(t,s)=\sum_{n=-\infty}^{+\infty}\lambda_n\varphi_n(t)\varphi_n(s),
$$
we see that all assumptions of Theorem~1 hold if we set
$$
\rho(u,t)=\cases{
0 &for $-1<u<1$,\cr
\sqrt{\frac{\lambda_n}{2h_n}}\varphi_n(t) &for $u\in[n,n+1)$,\cr
-\sqrt{\frac{\lambda_n}{2h_n}}\varphi_n(t) &for $u\in(-n-1,-n],\;\;\;
n=1,2,\dots.$ \cr}
$$

The solution of Eq.~(29) in the case of an absolutely continuous
measure $\nu$ was found for the Wiener measure, the conditional Wiener
measure, and some other measures. The following solution of~(29)
can be constructed for the Bogolyubov measure. We take the normalized
Lebesgue measure on the closed interval $[-\beta,\beta]$ as $\nu$, i.e.,
$$
d\nu(u)=\frac1{2\beta}du.
$$
Then
$$
\rho(u,t)=\sqrt{\frac\beta m}\,\frac1{e^{\beta\omega}-1}
e^{\omega(t-|u|)}\bigl[\theta\bigl(t-|u|\bigr)+
e^{\beta\omega}\theta\bigl(|u|-t\bigr)\bigr]\sign u.
\eqno(32)
$$
It can be verified that the measure $\nu$ and the function $\rho(u,t)$ thus
chosen satisfy all assumptions in Theorem~1. Hence, in the case of the
Bogolyubov measure in question, we have
$$
\int_XF(x)\,d\nn(x)\approx\frac1{(2\beta)^n}\int_{-\beta}^{\beta}
\dots\int_{-\beta}^{\beta}\,du_1\cdots du_n\,
F\biggl(\,\sum_{j=1}^nc_j\rho(u_j,t)\biggr)
$$
by Theorem~1, where $\rho(u,t)$ is given by~(32).

We consider approximate formulas that are exact for functional polynomials of
the third degree and for some functionals of a special form. As before, let
$X$ be the space $C[a,b]$ of continuous functions on $[a,b]$, let $\nu$ be a
measure on the Borel sets in the real line $R$, let $A(u)$ be a positive
function on $R$, and let $p(x)$ be a weight functional. We assume that the
conditions
$$
\int_RA(u)\,d\nu(u)\equiv A<\infty,
$$
$$
\int_Xp(x)x(t)\,d\mu(x)=\int_Xp(x)x(t)x(s)x(\tau)\,d\mu(x)\equiv0,
$$
$$
r(t,s)\equiv\frac1{p_0}\int_Xp(x)x(t)x(s)\,d\mu(x)<\infty
$$
hold, where
$$
p_0\equiv\int_Xp(x)\,d\mu(x).
$$

{\bf Theorem 3 {\rm\cite{25a2}}.}
{\sl
Let a symmetric function $r(t,s)$ be representable in the form
$$
r(t,s)=\int_Rx(u,t)x(u,s)\,d\nu(u),
$$
where the function $x(u,t)$ belongs to the space $L_2[R,\nu]$ relative to the
argument $u$. Then the formula
$$
\int_Xp(x)F(x)\,d\mu(x)\approx p_0(1-A)F(0)+
$$
$$
+\frac12p_0\int_RA(u)\biggl[F\biggl(\frac{x(u,\cdot)}{\sqrt{A(u)}}\biggr)+
F\biggl(-\frac{x(u,\cdot)}{\sqrt{A(u)}}\biggr)\biggr]\,d\nu(u)
\eqno(33)
$$
is exact if $F(x)$ is an arbitrary functional polynomial of the third degree.
}

If the measure $\nu$ is discrete, then
$$
A=\sum_kA_k,\qquad r(t,s)=\sum_kx_k(t)x_k(s),
$$
and formula~(33) becomes
$$
\int_Xp(x)F(x)\,d\mu(x)\approx p_0(1-A)F(0)+
\frac{p_0}2\sum_kA_k\left[F\left(\frac{x_k(\cdot)}{\sqrt{A_k}}\right)+
F\left(-\frac{x_k(\cdot)}{\sqrt{A_k}}\right)\right].
\eqno(34)
$$

We consider an example. Let the weight be
$$
p(x)=\int_a^bx^2(t)\,dt.
$$
Then
$$
r(t,s)=B(t,s)+\frac2{\tr B}\int_a^bB(t,\tau)B(\tau,s)\,d\tau,\qquad
\tr B=\int_a^bB(t,t)\,dt.
$$
If the expansion of the correlation function in a series with respect to its
eigenfunctions is used, then we obtain
$$
r(t,s)=\sum_k\left(\lambda_k+\frac2{\tr B}\lambda_k^2\right)
\varphi_k(t)\varphi_k(s),
$$
where $\lambda_k$ are the eigenvalues of the kernel $B(t,s)$.
Formula~(34) becomes
$$
\int_X\biggl(\,\int_a^bx^2(t)\,dt\biggr)F(x)\,d\mu(x)\approx
$$
$$
\approx\tr B\biggl\{(1-A)F(0)+\frac12\sum_kA_k\bigl[F\bigl(b_k
\varphi_k(\cdot)\bigr)+
F\bigl(-b_k\varphi_k(\cdot)\bigr)\bigr]\biggr\},
$$
where
$$
b_k=\left[\frac1{A_k}\left(\lambda_k+
\frac2{\tr B}\lambda_k^2\right)\right]^{1/2},\qquad
\sum_kA_k=A<\infty.
$$

{\bf Theorem 4 {\rm\cite{25a2}}.}
{\sl
Let the functions $r(t,s)$ and
$$
\rho(t,s)=\frac1{p_0}\int_Xp(x)V(x)x(t)x(s)\,d\mu(x),
$$
where $V(x)$ is an even positive functional on $C$, be representable in the
form
$$
r(t,s)=\sum_kx_k(t)x_k(s),\qquad\rho(t,s)=\sum_kB_kx_k(t)x_k(s),
$$
where $B_k$ are such that the equation
$$
V\left(\frac{x_k(\cdot)}{\sqrt{A_k}}\right)=B_k
$$
for each value of $k$ has a positive solution $A_k$ satisfying the condition
$$
A=\sum_kA_k<\infty.
$$
Then formula~(34) is exact for all functional polynomials of
the third degree and also for the functionals $F(x)$ of the form
$$
F(x)=V(x)p_2(x),
$$
where $p_2(x)$ is an arbitrary homogeneous functional of the second degree.
}

As an example, we consider the case of the Bogolyubov measure. Let
$$
p(x)\equiv1,\qquad V(x)=\|x\|^2=\int_0^{\beta}x^2(t)\,dt.
$$
Then
$$
\rho(t,s)=\sum_kB_kx_k(t)x_k(s),
$$
where
$$
B_k=\tr B+2\lambda_k,\qquad x_k(t)=\sqrt{\lambda_k}\,\varphi_k(t),
$$
and $\lambda_k$ and $\varphi_k(t)$ are the eigenvalues and
eigenfunctions of the kernel $B(t,s)$. In the case under consideration,
the other quantities in formula~(34) are given by the relations
$$
\tr B=\frac{\beta}{2m\omega}\coth\frac{\beta\omega}2,\qquad
A_k=\left(2+\frac\beta{2\omega}\coth\frac{\beta\omega}2
\bigl[\omega^2+(2\pi k\beta^{-1})^2\bigr]\right)^{-1},
$$
$$
A=\frac1{\sqrt{1+4\tanh(\beta\omega/2)/(\beta\omega)}\,}
\frac{\coth\left(\beta\omega\sqrt{1+4
\tanh(\beta\omega/2)/(\beta\omega)}\,/2\right)}{\coth(\beta\omega/2)}.
$$

\section { Stochastic processes and the Bogolyubov measure}

The notions and methods of the theory of stochastic processes are widely used
to study probability measures in function spaces. We assume that a
probability space $\{\Omega,G,P\}$ is fixed, where $\Omega$ is a space of
elementary events $\omega$ with a selected $\sigma$-algebra of subsets of
events $G$ and a measure, namely, the probability $P$ of events on $G$. The
numerical functions $f(\omega)$ on $\Omega$ measurable with respect to $P$
are called random variables. For integrable functions with respect to the
measure $P$, the integral (mathematical expectation)
$$
Mf(\omega)=\int_\Omega f(\omega)\,dP(\omega)
$$
is defined. By a random element with a range in $X$, we mean a weakly
measurable mapping $x(\omega)$ of $\Omega$ into $X$, i.e., a mapping under
which a functional $\br\xi,x(\omega)\kt$ is measurable with respect to the
measure $P$ for any $\xi\in X'$, where $X'$ is the adjoint space of $X$. If a
random element $x(\omega)$ with a range in $X$ is given, then the probability
measure
$$
\mu\bigl\{x\in X:[\br\xi_1,x\kt,\dots,\br\xi_n,x\kt]\in A_n\bigr\}=
P\bigl\{\omega\in\Omega:[\br\xi_1,x(\omega)\kt,\dots,
\br\xi_n,x(\omega)\kt]\in A_n\bigr\}
$$
can be defined on the $\sigma$-algebra generated by the cylindrical sets in
$X$. Here, $A_n$ is an arbitrary Borel set in $R^n$, and the vectors
$\xi_1,\dots,\xi_n$ $(n=1,2,\dots)$ belong to the adjoint space $X'$. In this
case,
$$
\int_\Omega F[x(\omega)]\,dP(\omega)=\int_X F(x)\,d\mu(x)
$$
for any functional $F$ such that one of the above integrals exists for it.

Let $X$ be a space of real functions of the argument $t\in T$, where $T$ is a
subset in $R$. Then a random element $x(\omega)$ is called a random function
and is denoted by $x(\omega,t)$. The argument $\omega$ in $x(\omega,t)$ is
often omitted. If $t$ is interpreted as time, then the related random
functions are called random or stochastic processes. A random function is
regarded as being defined if its finite-dimensional distributions are known.
A random function $x(t)=x(\omega,t)$ $(t\in T)$ with a range in a probability
space $\{X,G,P\}$ is called a Gaussian process if all its finite-dimensional
distributions are Gaussian. This means that the joint distribution of the
values $x(t_1),x(t_2),\dots,x(t_n)$ of this random process are defined by the
density function
$$
p(u_1,\dots,u_n)=(2\pi)^{-n/2}(\det B)^{-1/2}\exp\biggl[-\frac12
\sum_{i,j=1}^nb_{ij}^{(-1)}(u_i-m(t_i))(u_j-m(t_j))\biggr]
$$
with the mathematical expectation $m(t)=M[x(\omega,t)]$ and the correlation
function
$$
B(t,s)=M\bigl[(x(\omega,t)-m(t))(x(\omega,s)-m(s))\bigr],
$$
where $B$ is a matrix with the elements $B(t_i,t_j)$ $(i,j=1,2,\dots,n)$ and
$b_{ij}^{(-1)}$ are the elements of the matrix $B^{-1}$ inverse to $B$.
Therefore, if $X$ is a function space, then the relation
$$
\int_\Omega F[x(\omega,t)]\,dP(\omega)=\int_X F[x(t)]\,d\mu(x)
$$
holds, and the problem of integrating with respect to the Gaussian measure in
the function space is equivalent to the problem of integrating with respect
to the measure generated by the corresponding Gaussian random process. In
what follows, we constantly use this relation between the theory of Gaussian
random processes and the theory of functional integration with respect to
Gaussian measures.

Let $\vec t=(t_1,t_2,\dots,t_n)$, $0<t_1<t_2<\dots<t_n\le\beta$, be a set of
real numbers. For an arbitrary given subset $E\subset R^n$, we define the
cylindrical set $Q_{\vec t}\,(E)=\{x\in X:\ (x(t_1),\dots,x(t_n))\in E\}$.
The sets $E$ and $Q_{\vec t}(E)$ uniquely define each other for a fixed
$\vec t$. By definition, the centered Gaussian measure of the given
cylindrical set $Q_{\vec t}\,(E)$ is
$$
\mu\{Q_{\vec t}\,(E)\}=(2\pi)^{-n/2}(\det K)^{-1/2}\int_E
\exp\biggl(-\frac12\sum_{i,j=1}^nk_{ij}^{(-1)}u_iu_j\biggr)\,du_1\cdots du_n.
\eqno(35)
$$

In the case of the Bogolyubov measure~\cite{13a2}, we have
$X=C^\circ[0,\beta]$, where $C^\circ[0,\beta]$ is the space of continuous
functions on the closed interval $[0,\beta]$ with the uniform metric
$$
\rho=\max_{t\in[0,\beta]}|x(t)-y(t)|
$$
that satisfy the condition $x(0)=x(\beta)$. The bilinear functional
$K(\varphi,\psi)$ on the adjoint space $X'$ has the form
$$
K(\varphi,\psi)=\int_0^\beta\int_0^\beta B(t,s)\,d\varphi(t)\,d\psi(s),
$$
where $\varphi(t)\in X'=V_0[0,\beta]$ and $X'=V_0[0,\beta]$ is the space of
functions of bounded variation on $[0,\beta]$ satisfying the condition
$$
\varphi(0)=0,\qquad
\varphi(t)=\frac12[\varphi(t+0)+\varphi(t-0)]\quad\mbox{for }t\in(0,\beta).
$$
The elements of the variance matrix have the form~\cite{16a2}
$$
k_{ij}=B(t_i,t_j)=\frac1{2m\omega\sinh(\beta\omega/2)}
\cosh\left(\omega|t_i-t_j|-\frac{\beta\omega}2\right).
\eqno(36)
$$
The Bogolyubov measure has a zero mean.

We consider some special cases of formula~(35) for the Bogolyubov measure.

Let $0<t\le\beta$. We calculate the function
$$
F_{x(t)}\equiv\nn\{x\in X:x(t)\le\gamma\},
$$
where $\gamma$ is an arbitrary real number. Using~(35) and~(36), we
write
$$
F_{x(t)}=\frac1{\sqrt{2\pi
K(\varphi,\varphi)}}\int_{-\infty}^\gamma
\exp\left(-\frac12\frac{u^2}{K(\varphi,\varphi)}\right)\,du=
$$
$$
=\frac1{\sqrt{\bigl(\pi/(m\omega)\bigl)\coth(\beta\omega/2)}}
\int_{-\infty}^\gamma\exp\left(-\frac{u^2}2
\frac{2m\omega}{\coth(\beta\omega/2)}\right)\,du.
$$
This formula shows that the random variable $G(x)=x(t)$ is normally
distributed with a zero mean and the variance $(2m\omega)^{-1}
\coth(\beta\omega/2)$.

Let $0<t_1<t_2\le\beta$, and let $\gamma$ be an arbitrary real number. We
find the function
$$
F_{x(t_2)-x(t_1)}=\nn\left\{x\in X:x(t_2)-x(t_1)\le\gamma\right\}.
$$
We can write
$$
F_{x(t_2)-x(t_1)}=\nn\left\{x\in X:(x(t_1),x(t_2))\in E\right\},
$$
where $E=\{(u_1,u_2)\in R^2:u_2-u_1\le\gamma\}$. Using~(35), we obtain
$$
F_{x(t_2)-x(t_1)}=\frac1{2\pi}\frac1{\sqrt{\det K}}\int_Bdu_1\,du_2\,
\exp\biggl[-\frac12\bigl(k_{11}^{(-1)}u_1^2+k_{12}^{(-1)}u_1u_2+
$$
$$
+k_{21}^{(-1)}u_2u_1+k_{22}^{(-1)}u_2^2\bigr)\biggr].
\eqno(37)
$$
The elements of the inverse matrix $K^{-1}$ are calculated quite simply in
this case. They have the forms
$$
k_{11}^{(-1)}=k_{22}^{(-1)}=\frac{k_{11}}{\det K},\;\;
k_{12}^{(-1)}=k_{21}^{(-1)}=-\frac{k_{12}}{\det K},
$$
where
$$
k_{11}=\frac1{2m\omega\sinh(\beta\omega/2)}\cosh\frac{\beta\omega}2,\qquad
k_{12}=\frac1{2m\omega\sinh(\beta\omega/2)}
\cosh\left(\omega|t_1-t_2|-\frac{\beta\omega}2\right),
$$
$$
\det K=\frac1{4m^2\omega^2{\sinh}^2(\beta\omega/2)}
\left[\cosh^2\frac{\beta\omega}2-\cosh^2\left(\omega|t_1-t_2|-
\frac{\beta\omega}2\right)\right].
$$
After substituting these expressions in~(37) and performing some
elementary transformations, we obtain
$$
F_{x(t_2)-x(t_1)}={}\sqrt{
\frac{m\omega\sinh(\beta\omega/2)}{2\pi\left[\cosh(\beta\omega/2)-
\cosh(\beta\omega/2-\omega(t_2-t_1))\right]}}\,\times
$$
$$
\times\int_{-\infty}^\gamma du\,\exp\left[-\frac{u^2}2
\frac{m\omega\sinh(\beta\omega/2)}{\cosh(\beta\omega/2)-
\cosh(\beta\omega/2-\omega(t_2-t_1))}\right].
$$
Consequently, the random variable $G(x)=x(t_2)-x(t_1)$ is normally
distributed with a zero average and the variance
$$
\frac{\cosh(\beta\omega/2)-\cosh(\beta\omega/2-\omega(t_2-t_1))}
{m\omega\sinh(\beta\omega/2)}.
\eqno(38)
$$

\section { Metric properties of Bogolyubov trajectories }

\subsection{ Nondifferentiability of Bogolyubov trajectories}

We consider the properties of the support of the Bogolyubov measure in the
space $C^\circ[0,\beta]$. As in the case of the Wiener measure, the measure
$\nn$ is concentrated on continuous paths rather than on continuously
differentiable ones. Hence, along with the Wiener measure, the Bogolyubov
measure gives another important example of continuous functions that are
almost everywhere nondifferentiable.

We introduce the set
$$
C_h^\gamma(t,t')=\bigl\{x\in X:|x(t)-x(t')|\le h|t-t'|^\gamma\bigr\},
$$
where $h>0$, $0<\gamma\le1$, and $t,t'\in[0,\beta]$. We seek the Bogolyubov
measure of this set. Using~(35), we can write
$$
\nn\left\{C_h^\gamma(t,t')\right\}=\frac1{2\pi\sqrt{\det K}}
\int_B du_1du_2e^{-(u,K^{-1}u)/2},
\eqno(39)
$$
where $B=\bigl\{(u_1,u_2)\in R^2:|u_1-u_2|\le h|t-t'|^\gamma\bigr\}$. The
matrix $K$ in~(39) coincides with the matrix $K$ used in the
preceding section in formula~(37). Performing a suitable linear
change of integration variables in~(39), we obtain
$$
\nn\left\{C_h^\gamma(t,t')\right\}=
\frac1{\sqrt{2\pi}}\int_{-a}^ae^{-v^2/2}\,dv,
\eqno(40)
$$
where
$$
a=\sqrt{\frac{m\omega\sinh(\beta\omega/2)}{\cosh(\beta\omega/2)-
\cosh(\beta\omega/2-\omega|t-t'|)}}\,h\left|t-t'\right|^\gamma.
$$
Formula~(40) implies an upper estimate for the desired measure,
$$
\nn\left\{C_h^\gamma(t,t')\right\}\le\sqrt{\frac2{\pi}}a.
\eqno(41)
$$

We now consider the sets
$$
C_h^\gamma(t)=\bigcap_{t'\in[0,\beta]}C_h^\gamma(t,t')=\bigl\{x\in X:
|x(t)-x(t')|\le h|t-t'|^\gamma\mbox{ for all }t'\in[0,\beta]\bigr\}
\eqno(42)
$$
and
$$
C_h^\gamma=\bigcap_{t\in[0,\beta]}C_h^\gamma(t)=\bigl\{x\in X:
|x(t)-x(t')|\le h|t-t'|^\gamma\mbox{ for all }t,t'\in[0,\beta]\bigr\}.
$$
It can be proved~\cite{3a3} that $C_h^\gamma(t,t')$, $C_h^\gamma(t)$ , and
$C_h^\gamma$ are closed subsets in $X=C^\circ[0,\beta]$.

We consider a sequence of points $\{t_k\}$ in the closed interval $[0,\beta]$
such that they do not coincide with $t$ and $t_k\to t$ as $k\to\infty$. Then
definition~(42), inequality~(41), and the downward
convexity of $\cosh x$ imply that
$$
\nn\left\{C_h^\gamma(t)\right\}\le\nn\left\{C_h^\gamma(t,t_k)\right\}\le
\sqrt{\frac{2m\sinh(\beta\omega/2)}
{\pi
\sinh(\beta\omega/2-\omega|t-t_k|)}}\,h\left|t-t_k\right|^{\gamma-1/2}.
$$
The resulting inequality shows that $\nn\left\{C_h^\gamma(t)\right\}=0$ for
$\gamma>1/2$, and consequently
$$
\nn\left\{C_h^\gamma\right\}=0.
\eqno(43)
$$

We recall that a function $x\:[0,\beta]\mapsto R$ is said to be H\"older
continuous of order $\gamma$ if there is a positive constant $h$ such that
$|x(t)-x(t')|\le h|t-t'|^\gamma$ for all $t,t'\in[0,\beta]$ and
$\gamma\in(0,t]$. Because
$$
\Gamma^\gamma\equiv\bigl\{x\in X:x\mbox{ is a H\"older continuous function
of order }\gamma\bigr\}=\bigcup_{h=1}^\infty C_h^\gamma,
$$
condition~(43) implies that $\Gamma^\gamma$, $1/2<\gamma\le1$,
is a Borel subset in $X$ with the Bogolyubov measure (or probability)
zero.  In other words, the Bogolyubov trajectories are not H\"older
continuous of order $\gamma>1/2$ almost everywhere with respect to the
measure.  (Consequently, they cannot be continuously differentiable.)

Let $0\le t\le\beta$. We consider the set $D_t=\{x\in X:x'(t)
\mbox{ exists}\}$, where $x'(t)$ denotes the ordinary derivative of $x$ with
respect to $t$ for $t\in(0,\beta)$ and the one-sided derivative for $t=0$ or
$t=\beta$. It can then be shown~\cite{3a3} that
$D_t\subset\bigcup_{h=1}^\infty C_h^1(t)$, whence $\nn(D_t)=0$.

We define a function $F=X\times[0,\beta]\mapsto R$ by the relation
$$
F(x,t)=\cases{
1, & if $x'(t)$ exists (as a finite function),\cr
0  & otherwise.\cr}
$$
It can be proved~\cite{3a3} that $F$ is measurable as a function of $x$ and
$t$. Therefore, by the Fubini theorem,
$$
\int_X\biggl(\,\int_0^\beta F(x,t)\,dt\biggr)\,d\nn(x)=
\int_0^\beta\biggl(\,\int_X F(x,t)\,d\nn(x)\biggr)\,dt=
\int_0^\beta\nn(D_t)\,dt=0.
$$
This formula shows that the relation
$$
\int_0^\beta F(x,t)\,dt=0
$$
holds for almost all functions $x$ with respect to the measure $\nn$.
Consequently, the relation $F(x,t)=0$ holds for almost all functions $x$ with
respect to the Bogolyubov measure $\nn$ and for almost all values of $t$ with
respect to the Lebesgue measure. We have thus proved that the trajectories
$x\in X$ are differentiable with probability 1 on at most a subset in
$[0,\beta]$ of Lebesgue measure zero.

Because every function $x$ of bounded variation on any interval is always
everywhere differentiable with respect to the Lebesgue measure on this
interval~\cite{4a3}, the Bogolyubov trajectories have unbounded variation
with probability 1 on any subinterval of $[0,\beta]$.

\subsection { Scale transformations in the Bogolyubov space}

In the theory of Feynman continual integrals, the scale transformation
$x\mapsto\sigma x$ with the parameter $\sigma\in C$ in the related function
space is important. An essential role is played here by the well-known L\'evy
theorem on the quadratic variation of Wiener trajectories~\cite{5a3} and
by the special case that was investigated somewhat later in~\cite{6a3}. In
view of the possible analytic continuation with respect to temperature or
mass in the Bogolyubov continual integral, it is interesting to apply the
L\'evy scheme to the case of the Bogolyubov measure.

We introduce L\'evy quadratic variations of trajectories. We consider a
partition $\Pi$ of the closed interval $[0,\beta]$, $0=t_0<t_1<\dots<t_k=
\beta$, and a function $x\in X$. We define the function
$$
S_\Pi(x)=\sum_{j=1}^k\bigl[x(t_j)-x(t_{j-1})\bigr]^2.
\eqno(44)
$$
If the interval $[0,\beta]$ is partitioned into $k$ equal subintervals, then
we simply write $S_k(x)$ instead of $S_\Pi(x)$. We note that
$$
\lim_{n\to\infty} S_{2^n}(x)=0
\eqno(45)
$$
for sufficiently smooth trajectories, for example, for those satisfying the
Lipschitz condition with constant $k$. However, as shown in the preceding
section, the Bogolyubov measure is concentrated on nondifferentiable
trajectories. Therefore, as in the case of the Wiener measure, it can be
expected that condition~(45) does not hold for Bogolyubov
trajectories.

{\bf Theorem 5.}
{\sl The Bogolyubov trajectories $x\in X$ satisfy the relation
$$
\lim_{n\to\infty}S_{2^n}(x)=\frac\beta m
$$
almost everywhere.
}

{\bf Proof.}
We first show that
$$
I_N\equiv\left\|S_N-\frac\beta
m\right\|_2^2=\frac{2\beta^2}{m^2}\frac1N+O\left(\frac1{N^2}\right)
\eqno(46)
$$
for any sufficiently large positive integer $N$, where $\|\,\cdot\,\|_2\equiv
\|\,\cdot\,\|_{L^2(X,\nn)}$ is the $L^2$-norm in the space $X$ with the
measure $\nn$. It follows from definition~(44) that
$$
S_N(x)=\sum_{j=1}^N\left[x(t_j)-x(t_{j-1})\right]^2,\qquad
t_j=\frac{j\beta}N,\quad j=0,1,\dots,N.
$$
We have
$$
I_N=\int_X\left(S_N(x)-\frac\beta m\right)^2\,d\nn(x)=
$$
$$
=\int_XS_N^2(x)\,d\nn(x)-\frac{2\beta}m\int_X S_N(x)\,d\nn(x)+
\frac{\beta^2}{m^2}
\eqno(47)
$$
for the desired expression $I_N$. The random variable
$x(t_j)-x(t_{j-1})$ is distributed with a zero mean and variance~(38),
and therefore
$$
\int_X S_N(x)\,d\nn(x)=\sum_{j=1}^N\frac{\cosh(\beta\omega/2)-
\cosh(\beta\omega/2-\omega|t_j-t_{j-1}|)}{m\omega\sinh(\beta\omega/2)}=
$$
$$
=N\frac{\cosh(\beta\omega/2)-\cosh(\beta\omega/2-\beta\omega/N)}
{m\omega\sinh(\beta\omega/2)}.
$$
In particular,
$$
\int_XS_N(x)\,d\nn(x)=\frac\beta m-\frac{\beta^2\omega}{2m}
\coth\frac{\beta\omega}2\frac1N+O\left(\frac1{N^2}\right)
$$
as $N\to\infty$. We now calculate the integral of $S_N^2(x)$ in
formula~(47),
$$
\int_XS_N^2(x)\,d\nn(x)={}\sum_{m,n=1}^N\int_X\,d\nn(x)\bigl[
x^2(t_n)x^2(t_m)+2x^2(t_n)x^2(t_{m-1})+
$$
$$
+x^2(t_{n-1})x^2(t_{m-1})-4x^2(t_n)x(t_m)x(t_{m-1})-
$$
$$
-4x^2(t_{n-1})x(t_m)x(t_{m-1})+4x(t_n)x(t_{n-1})x(t_m)x(t_{m-1})\bigr].
\eqno(48)
$$
In calculating the integrals of individual terms in the right-hand side
of~(48), it is necessary to use the Wick theorem and the corresponding
tabular values of finite sums in~\cite{7a3}. For example,
$$
\int_X d\nn(x)\,x^2(t_n)x^2(t_m)=B^2(t_n,t_n)+2B^2(t_n,t_m).
$$
Using~(36), we find
$$
\sum_{n,m=1}^N \int_X d\nn(x)\,x^2(t_n)x^2(t_m)=
\left(2m\omega\sinh\frac{\beta\omega}2\right)^{-2}\times
$$
$$
\times\left(N^2\cosh^2\frac{\beta\omega}2+N^2+
N\sinh(\beta\omega)\coth\frac{\beta\omega}N\right).
$$
Accordingly, calculating the other terms in the right-hand side
of~(48), we obtain
$$
\int_X S_N^2(x)\,d\nn(x)=
\left(2m\omega\sinh\frac{\beta\omega}2\right)^{-2}\Biggl[
4N^2\cosh^2\frac{\beta\omega}2+4N^2\cosh^2\left(\frac{\beta\omega}2-
\frac{\beta\omega}N\right)-
$$
$$
-8N^2\cosh\frac{\beta\omega}2\cosh\left(\frac{\beta\omega}2-
\frac{\beta\omega}N\right)+6N^2-8N^2\cosh\frac{\beta\omega}N+
$$
$$
+2N(N-1)\cosh\frac{2\beta\omega}N+2N+
2N\cosh\left(\beta\omega-\frac{2\beta\omega}N\right)+
$$
$$
+6N\sinh(\beta\omega)\frac{\cosh(\beta\omega/N)}{\sinh(\beta\omega/N)}-
8N\frac{\sinh(\beta\omega)}{\sinh(\beta\omega/N)}+2N
\frac{\sinh\bigl(\beta\omega-(\beta\omega/N)\bigr)}
{\sinh(\beta\omega/N)}\Biggr].\quad
\eqno(49)
$$
With regard to passage to the limit as $N\to\infty$, we obtain
$$
\int_XS_N^2(x)\,d\nn(x)=\frac{\beta^2}{m^2}+
\frac{\beta^2}{2m^2}\frac{2\cosh(\beta\omega)-\beta\omega\sinh(\beta\omega)-2}
{\sinh^2(\beta\omega/2)}\,\frac1N+\vep_N
$$
from formula~(49), where $\vep_N\sim O(1/N^2)$, is a positive
number.  As a result, we obtain relation~(46) for the desired
expression $I_N$.  In particular, it follows from~(46) that
$$
\left\|S_{2^n}-\frac\beta m\right\|_2^2=
\frac{\beta^2}{m^2}\frac1{2^{n-1}}+\vep_{2^n}.
\eqno(50)
$$
We consider the set
$$
E_n=\left\{x\in X:\left|S_{2^n}(x)-\frac\beta m\right|\ge
\frac\beta m\frac1{2^{n/3}}+\mu_n\right\},
\eqno(51)
$$
where
$$
\mu_n=\frac\beta m\frac1{2^{n/3}}\left(\sqrt{1+\frac{m^2}{\beta^2}2^{n-1}
\vep_n}-1\right)\sim O\left(\frac1{2^{4n/3+1}}\right).
$$
We prove that
$$
\nn(E_n)\le\frac2{2^{n/3}}.
\eqno(52)
$$
We suppose the contrary. Then
$$
\int_X\left|S_{2^n}(x)-\frac\beta m\right|^2\,d\nn(x)\ge
\int_{E_n}\left|S_{2^n}(x)-\frac\beta m\right|^2d\nn(x)>
$$
$$
>\left(\frac\beta m\frac1{2^{n/3}}+\mu_n\right)^2\frac2{2^{n/3}}=
\frac{\beta^2}{m^2}\frac1{2^{n-1}}+\vep_{2^n},
$$
which contradicts~(50). We set
$$
F_n=\bigcup_{k=n}^\infty E_k.
$$
Then it follows from~(52) that
$$
\nn(F_n)\le\sum_{k=n}^\infty\nn(E_k)\le\frac c{2^{n/3}},
\eqno(53)
$$
where $c=2^{4/3}(2^{1/3}-1)^{-1}$. Formula~(51) now implies
that the inequality
$$
\left|S_{2^k}(x)-\frac\beta
m\right|<\frac\beta m2^{-k/3}
$$
holds for $x\in X\setminus
F_n=\bigcap_{k=n}^\infty E_k^{\rm{c}}$, where $E_k^{\rm{c}}$ is the
complement of the set $E_k$ in $X$, and for any $k=n,n+1,\dots$\,.
Consequently, if there is an $n$ such that $x\notin F_n$, then
$$
\lim_{k\to\infty}S_{2^k}(x)=\frac\beta m.
\eqno(54)
$$
Therefore, condition~(54) holds for all $x$ possibly except for
$x\in F=\bigcap_{n=1}^\infty F_n$. However, inequality~(53)
implies that
$$
\nn(F)\le\nn(F_n)\le\frac c{2^{n/3}}
$$
for any $n$,
i.e., $\nn(F)=0$. Theorem 5 is proved.

We consider the set
$$
\Omega_\sigma=\left\{x\in X:\lim_{n\to\infty}S_{2^n}(x)=\sigma^2
\frac\beta m\right\},
$$
where $\sigma$ is a given positive number. Let $\nn^\sigma$ $(\sigma>0)$
denote the image of the measure $\nn\equiv\nn^1$ under the mapping
$\varphi_\sigma\:X\mapsto X$, $\varphi_\sigma=\sigma x$. The measure
$\nn^\sigma=\nn^1\circ\sigma^{-1}$ is defined on the Borel $\sigma$-algebra
$\cal B(X)$, and the relation $\nn^\sigma(B)=\nn^1(\sigma^{-1}B)$ holds for
any $B\in\cal B$.

{\bf Proposition 1~\cite{3a3}.}

{\sl
$1.$ The set $\Omega_\sigma$ is Borel measurable for any $\sigma>0$.

$2.$ The relation $\sigma_2\Omega_{\sigma_1}=\Omega_{\sigma_1\sigma_2}$ holds
for any $\sigma_1,\sigma_2>0$; in particular, $\sigma\Omega_1=\Omega_\sigma$
for any $\sigma>0$.

$3.$ For any $\sigma>0$, $\nn^\sigma(\Omega_\sigma)=1$.

$4.$ If $\sigma_1\ne\sigma_2$ $(\sigma_1,\sigma_2>0)$, then
$\Omega_{\sigma_1}\cap\Omega_{\sigma_2}=\emptyset$, i.e., the measures
$\nn^{\sigma_1}$ and $\nn^{\sigma_2}$ are mutually orthogonal.
}

Assertion~3 in the proposition implies that $\Omega_\sigma$ is a set of full
$\nn^\sigma$-measure. Following~\cite{3a3}, we say that the measure
$\nn^\sigma$ is concentrated on the set $\Omega_\sigma$. We note that
$\supp\nn^\sigma=X$ for any $\sigma>0$ and $\Omega_\sigma\subset X$.

A subset $A$ in $X$ is called a {\sl scale-invariant measurable set\/} if
$\sigma A\in\cal S_1$ for all $\sigma>0$, where $\cal S_1$ is the
$\sigma$-algebra of sets in the space $X$ that are measurable with respect to
the Bogolyubov measure $\nn^1$. A scale-invariant measurable set $N$ is
called a {\sl zero-scale-invariant measurable set\/} if $\nn^1(\sigma N)=0$
for all $\sigma>0$. The classes of scale-invariant and zero-scale-invariant
sets are denoted by $\cal S$ and $\cal N$ respectively. We let
$\cal S_\sigma$ denote the $\sigma$-algebra obtained by completing the
measurable space $(X,\cal B(X),\nn^\sigma)$ and $\cal N_\sigma$ denote
the class of $\nn^\sigma$-zero sets. It can be shown that $\cal N_\sigma=
\sigma\cal N_1$, $\cal S_\sigma=\sigma \cal S_1$, and
$\nn^\sigma(E)=\nn^1(\sigma^{-1}E)$ for any $E\in\cal S_\sigma$. Moreover,
the algebra $\cal S$ is a $\sigma$-algebra, $\cal S=\bigcap_{\sigma>0}
\cal S_\sigma$, and $\cal N=\bigcap_{\sigma>0} \cal N_\sigma$. It can be
easily seen that $\cal B(X)\subset\cal S\subset\cal S_\sigma$ for each
$\sigma>0$. We have $E\in\cal S$ if and only if $E\cap\Omega_\sigma\in
\cal S_\sigma$ for any $\sigma>0$, and $N\in\cal N$ if and only if
$N\cap\Omega_\sigma\in\cal N_\sigma$ for any $\sigma>0$.

The structure of scale-invariant and zero-scale-invariant sets is described
in the following proposition.

{\bf Proposition 2~\cite{3a3}.}

{\sl
$1.$ The inclusion $E\in \cal S$ holds if and only if the set $E$ can be
represented in the form
$$
E=\biggl(\,\bigcup_{\sigma>0}E_\sigma\biggr)\cup L,
\eqno(55)
$$
where each $E_\sigma$ is an $\nn^\sigma$-measurable subset in $\Omega_\sigma$
and $L$ is an arbitrary subset of the set
$$
X\Big\backslash\bigcup_{\sigma>0}\Omega_\sigma.
\eqno(56)
$$
Relation $\nn^\sigma(E)=\nn^\sigma(E_\sigma)$ holds for all $\sigma>0$ and
any set $E$ of form~(55).

$2.$ The inclusion $N\in\cal N$ holds if and only if the set $N$ can be
represented in the form
$$
N=\biggl(\,\bigcup_{\sigma>0}N_\sigma\biggr)\cup L,
$$
where each set $N_\sigma$ is an $\nn^\sigma$-measurable subset in
$\Omega_\sigma$ and $L$ is an arbitrary subset of the set
in~(56).
}

\section {Dynamic properties of Bogolyubov measure}

\subsection {Semigroups with respect to the Bogolyubov measure}

Let $\cal L(X)$ denote the space of bounded linear operators in a Banach
space $X$. We recall that a family of operators $\{T(t):0\le t<\infty\}$ in
$\cal L(X)$ is called a strongly continuous semigroup of operators on $X$
if $T(0)=I$, $T(t,s)=T(t)T(s)$ for all $t,s\in[0,\infty)$, and the mapping
$t\mapsto T(t)x$ from $[0,\infty)$ into $X$ is continuous for each $x\in X$.
As is known~\cite{8a3}, in the case of a strongly continuous semigroup, the
generating operator (generator)
$$
L=\lim_{\epsilon\to0}\frac1\epsilon(T(\epsilon)-I)
$$
of the semigroup has a dense domain $D(L)$ in $X$, is a closed linear
operator, and
$$
\lim_{\epsilon\to0}\frac1\epsilon\bigl(T(t+\epsilon)-T(t)\bigr)f=
LT(t)f=T(t)Lf
$$
for $f\in D(L)$. For a strongly continuous semigroup $\{T(t):0\le t<\infty\}$
with the generator $L$ and an arbitrary vector $f\in D(L)$, it can be
shown~\cite{9a3} that the function $u(t)=T(t)f\in X$ is continuously
differentiable on the half-infinite interval $[0,\infty)$ and satisfies the
initial condition $u(0)=f$ and that the differential equation $du/dt=Lu$ is
satisfied for all $t>0$.

In the case of Gaussian measures, there is a universal example of a strongly
continuous semigroup known as the Ornstein--Uhlenbeck semigroup. Let $\mu$ be
a centered Gaussian measure on a locally convex space $X$. The
Ornstein--Uhlenbeck semigroup on the space $L^p(\mu)$ is given by the formula
$$
T(t)f(x)=\int_Xf\bigl(e^{-t}x+\sqrt{1-e^{2t}}\,y\bigr)\,d\mu(y).
\eqno(57)
$$
It was proved~\cite{10a3} that for every $p\ge1$, the family of operators
$\{T(t):0\le t<\infty\}$ defined by formula~(57) forms a
strongly continuous semigroup on the Banach space $L^p(\mu)$ with the
operator norm
$$
\|T(t)\|_{\cal L(L^p(\mu))}=1.
$$
Moreover, the operators $T(t)$ are nonnegative for $p=2$.

In the case of the Bogolyubov measure, the form of the generator of the
Ornstein--Uhlenbeck semigroup can be found under the assumption that
$f\in C_0^\infty(R)$, where $C_0^\infty(R)$ is the space of infinitely
differentiable functions compactly supported in $R$. We have
$$
T(t)f(x)=\frac1{\sqrt{\pi\coth(\beta\omega/2)/(m\omega)}}
\int_{-\infty}^\infty f\bigl(e^{-t}x+\sqrt{1-e^{-2t}}\,y\bigr)
\exp\left(-\frac{m\omega y^2}{\coth(\beta\omega/2)}\right)\,dy.
$$
The change of the integration variable $e^{-t}x+\sqrt{1-e^{-2t}}\,y=z$
results in
$$
\frac{T(t)f(x)-f(x)}t=\frac1t\Biggl\{
\sqrt{\frac{m\omega\tanh(\beta\omega/2)}{\pi(1-e^{-2t})}}
\int_{-\infty}^\infty (f(z)-f(x))\times
$$
$$
\times\exp\biggl[-\frac{m\omega\tanh(\beta\omega/2)}
{1-e^{-2t}}(z-e^{-t}x)^2\biggr]\,dz\Biggr\}.
$$
We now use the Taylor theorem to expand $f(z)$ under the integral sign,
$$
f(z)=f(x)+f'(x)(z-x)+\frac12f''(x)(z-x)^2+\frac16f^{(3)}(x)(z-x)^3+
\frac{f^{(4)}(\xi)}{24}(z-x)^4.
$$
Furthermore, calculating the elementary Gaussian integrals, we obtain
$$
\frac{T(t)f(x)-f(x)}t=-xf'(x)+\frac12f''(x)\frac1{2t}
\frac{1-e^{-2t}}{m\omega\tanh(\beta\omega/2)}+o(t),
$$
i.e., the generator of semigroup~(57) in the case of the Bogolyubov
measure is given by
$$
L=-x\frac d{dx}+\frac{\coth(\beta\omega/2)}{2m\omega}\frac{d^2}{dx^2}.
$$

We next consider the family of operators $\{T(\beta):0\le\beta<\infty\}$
acting in the space $L^2(R)$ according to the formula
$$
(T(\beta)f)(x)=\int_X d\nn(y)\,f\biggl(\,\int_0^\beta y(t)\,dt+x\biggr).
\eqno(58)
$$
It is clear that $T(0)=I$. Moreover, using the formulas for integration with
respect to Gaussian measures~\cite{11a3}, we obtain
$$
(T(\beta)f)(x)=\sqrt{\frac{m\omega^2}{2\pi\beta}}\int_{-\infty}^\infty
f(y)\exp\left[-\frac{(y-x)^2m\omega^2}{2\beta}\right]\,dy.
\eqno(59)
$$
Formula~(59) gives the well-known expression for the free semigroup in
the case of the heat conduction equation. Hence, family of
operators~(58) is in fact a strongly continuous semigroup in $L^2(R)$.
In this case, the generator of the semigroup has the form
$$
L=\frac1{2m\omega^2}\frac{d^2}{dx^2},
$$
and for any $f\in L^2(R)$, the function $u(\beta,x)=(T(\beta)f)(x)$ is the
solution of the Bloch equation
$$
\frac{\ptl u}{\ptl\beta}
=\frac1{2m\omega^2}\frac{\ptl^2u}{\ptl x^2}
$$
with the initial condition $u(0,x)=f(x)$. Formula~(58) implies the
relation between the Bogolyubov and Wiener measures
$$
\int_{C^\circ[0,m\omega^2t]}f
\biggl(x+\int_0^{m\omega^2t}y(\tau)\,d\tau\biggr)\,d\nn(y)=
\int_{C_0^t}f(y(t)+x)\,d\mu_{\rm{W}}^{}(y),
$$
where $C_0^t$ is the space of continuous functions on $[0,t]$ vanishing at
zero.

\subsection { Independent increments}

The classical Wiener process on the interval $[a,b]$ has independent
increments, i.e., for any $a<t_1<t_2<\dots<t_n\le b$, the random variables
$\xi_{t_2}-\xi_{t_1},\dots,\xi_{t_n}-\xi_{t_{n-1}}$ are independent. To prove
this assertion, because the Wiener process is Gaussian, it suffices to show
that these increments are pairwise independent. In the case of the Bogolyubov
measure, the increments $x(t_i)-x(t_{i-1})$, $i=1,2,\dots,n$, are not
independent, which substantially hampers an analysis of the corresponding
random process. However, the relation between the Wiener and Bogolyubov
measures established in Subsec.~8.1 permits constructing a system of
independent increments for the Bogolyubov random process as well.

We consider the random variable
$$
y(t)=\lambda x(t)+\int_0^t x(\tau)\,d\tau,\quad 0\le t\le\beta,
$$
where $\lambda$ is a constant to be defined below. The mathematical
expectation $M(y(t)y(s))$ is given by
$$
M(y(t)y(s))=\left[\frac{\lambda^2}{2m\omega\sinh(\beta\omega/2)}-
\frac1{2m\omega^3\sinh(\beta\omega/2)}\right]
\cosh\left(\omega|t-s|-\frac{\beta\omega}2\right)+{}
$$
$$
+\frac1{2m\omega^2\sinh(\beta\omega/2)}\biggl[
2\sinh\frac{\beta\omega}2\min(s,t)-\frac1\omega\cosh\frac{\beta\omega}2+{}
$$
$$
+\frac1\omega\cosh\left(\omega s-\frac{\beta\omega}2\right)+
\frac1\omega\cosh\left(\omega t-\frac{\beta\omega}2\right)\biggr]+{}
$$
$$
+\frac\lambda{2m\omega^2\sinh(\beta\omega/2)}\left[
2\sinh\frac{\beta\omega}2+\sinh\left(\omega s-\frac{\beta\omega}2\right)+
\sinh\left(\omega t-\frac{\beta\omega}2\right)\right].
$$
Setting $\lambda=\omega^{-1}$, we can now easily show that
$$
M\bigl[(y(t)-y(s))(y(\tau)-y(\sigma))\bigr]=0\quad
\mbox{for }s<t<\sigma<\tau.
$$
Because the Bogolyubov process is Gaussian, we can state the above result in
the form of the following theorem.

{\bf Theorem 6.}
{\sl
A Gaussian random process with a Bogolyubov measure has independent
increments, i.e., the random variables $y(t_2)-y(t_1),\dots,
y(t_n)-y(t_{n-1})$, where
$$
y(t)=\omega^{-1}x(t)+\int_0^tx(\tau)\,d\tau,\quad 0\le t\le\beta,
\eqno(60)
$$
are independent for any $0<t_1<t_2<\dots<t_n\le\beta$.
}

The random process $\{y(t)$, $0\le t\le\beta\}$ is a Gaussian process with a
zero average and the correlation function
$$
M(y(t)y(s))={}\frac1{2m\omega^2\sinh(\beta\omega/2)}\Biggl\{
2\left[\omega^{-1}+\min(s,t)\right]\sinh\frac{\beta\omega}2-
\frac1\omega\cosh\frac{\beta\omega}2+{}
$$
$$
+\frac1\omega\left[\cosh\left(\omega s-\frac{\beta\omega}2\right)+
\sinh\left(\omega s-\frac{\beta\omega}2\right)\right]+{}
$$
$$
+\frac1\omega\left[\cosh\left(\omega t-\frac{\beta\omega}2\right)+
\sinh\left(\omega t-\frac{\beta\omega}2\right)\right]\Biggr\}.
\eqno(61)
$$
Formula~(61) permits proving that the Gaussian random variable
$G\equiv y(t)-y(s)$ is normally distributed with a zero mean and the variance
$(t-s)/(m\omega^2)$, where $t>s$, i.e.,
$$
G\sim N\left(0,\frac{t-s}{m\omega^2}\right).
$$

We note that if $x(t)$ is regarded as a random function, then the integral
$$
\int_0^tx(\tau)\,d\tau
\eqno(62)
$$
introduced in previous sections is a stochastic integral defined as the limit in
the mean with respect to the given measure for the corresponding integral
sums. Integral~(62) exists if and only if the mean value $M(y^2)$
exists. This condition is fulfilled for the Bogolyubov measure, which, in
particular, follows from formula~(61).

To conclude this subsection, we note that because
$$
\int_Xd\nn(x)\biggl(\frac1\beta\int_0^\beta x(t)\,dt\biggr)^2=
\frac1{\beta^2}\int_0^\beta dt\int_0^\beta d\tau\,B(t-\tau)=
\frac1{\beta m\omega^2},
$$
we have
$$
\lim_{\beta\to\infty}M\biggl(\frac1\beta\int_0^\beta x(t)\,dt\biggr)^2=0.
$$
Because the Bogolyubov random process has a zero mathematical expectation,
$m\equiv Mx(t)=0$, we can say that this is an ergodic process in the sense
that the ``temporal" means (with respect to $\beta$) converge in the squared
mean to the ``phase" means.

\subsection { Bogolyubov measure and differential equations}

We define the function
$$
\delta_{\beta,\xi}(x)=
\frac1{2\pi}\int_{-\infty}^\infty dz\,e^{iz[y(\beta)-y(0)-\xi]},
\eqno(63)
$$
where $x(t)\in X$ is an arbitrary function, $\xi$ is an arbitrary real
number, $\beta$ is a positive number, and $y(t)$, $0\le t\le\beta$, is
defined in~(60). Function~(63) is an analogue of the
Donsker--Lions function~\cite{12a3}, which was introduced some time ago to
investigate the Wiener measure.

{\bf Lemma.}
{\sl
The mathematical expectation of function~(63) is given by
$$
E_{\nn}\left\{\delta_{\beta,\xi}(x)\right\}=
\sqrt{\frac{m\omega^2}{2\pi\beta}}
\exp\left(-\frac{m\omega^2}{2\beta}\xi^2\right).
\eqno(64)
$$
}

{\bf Proof.}
We consider the mathematical expectation
$$
E_{\nn}\left\{\delta_{\beta,\xi}(x)\right\}=
\int_X\delta_{\beta,\xi}(x)\,d\nn(x)=
$$
$$
=\frac1{2\pi}\int_{-\infty}^\infty dz\,e^{-iz\xi}
\int_Xd\nn(x)\,\exp\biggl[iz\int_0^\beta x(t)\,dt\biggr]=
$$
$$
=\frac1{2\pi}\int_{-\infty}^\infty dz\,e^{-iz\xi}\frac1{\sqrt{2\pi a}}
\int_{-\infty}^\infty e^{izu}e^{-u^2/(2a)}\,du,
\eqno(65)
$$
where
$$
a=\int_0^\beta\int_0^\beta B(t,s)\,dt\,ds=\frac\beta{m\omega^2}.
$$
Calculating the integrals in~(65), we derive~(64). The lemma is
proved.

We introduce the function
$$
u(\beta,\xi)=E_{\nn}\biggl\{\delta_{\beta,\xi}(x)
\exp\biggl(-\int_0^\beta V(y(s)-y(0))\,ds\biggr)\biggr\},
\eqno(66)
$$
where $V$ is a real function bounded from below.

{\bf Theorem 7.}
{\sl
Function~(66) is a solution of the partial differential equation
$$
\frac{\ptl u}{\ptl\beta}=\frac1{2m\omega^2}\frac{\ptl^2u}
{\ptl \xi^2}-V(\xi)u
\eqno(67)
$$
with the initial condition $u(0,\xi)=\delta(\xi)$ and the boundary conditions
$u(\beta,\pm\infty)=0$.
}

{\bf Proof.}
We use the obvious formula
$$
\exp\biggl(-\int_0^t V(z(s))\,ds\biggr)
=1-\int_0^tV(z(\tau))\exp\biggl(-\int_0^\tau V(z(s))\,ds\biggr)\,d\tau.
$$
This gives
$$
u(\beta,\xi)=E_{\nn}\left\{\delta_{\beta,\xi}(x)\right\}
-\int_0^\beta E_{\nn}\biggl\{\delta_{\beta,\xi}(x)V(y(\tau)-y(0))
\exp\biggl(-\int_0^\tau V(y(s)-y(0))\,ds\biggr)\biggr\}\,d\tau=
$$
$$
=E_{\nn}\left\{\delta_{\beta,\xi}(x)\right\}-\frac1{2\pi}
\int_0^\beta d\tau\int_{-\infty}^\infty dz\,e^{-iz\xi}\times{}
$$
$$
\times E_{\nn}\biggl\{V(y(\tau)-y(0))
\exp\biggl(-\int_0^\tau V(y(s)-y(0))\,ds+iz(y(\beta)-y(0))\biggr)\biggr\}.
$$
At the same time, we have
$$
E_{\nn}\biggl\{V(y(\tau)-y(0))
\exp\biggl(-\int_0^\tau V(y(s)-y(0))\,ds+iz(y(\beta)-y(0))\biggr)\biggr\}=
$$
$$
=E_{\nn}\biggl\{\biggl[V(y(\tau)-y(0))
\exp\biggl(-\int_0^\tau V(y(s)-y(0))\,ds+
iz(y(\tau)-y(0))\biggr)\biggr]\times{}
$$
$$
\times\bigl[\exp\bigl({iz(y(\beta)-y(0))-iz(y(\tau)-y(0))\bigr)}
\bigr]\biggr\}=
$$
$$
=E_{\nn}\biggl\{V(y(\tau)-y(0))
\exp\biggl(-\int_0^\tau V(y(s)-y(0))\,ds+
iz(y(\tau)-y(0))\biggr)\biggr\}\times{}
$$
$$
\times E_{\nn}\bigl\{\exp\bigl(iz(y(\beta)-y(\tau))\bigr)\bigr\}=
$$
$$
=\exp\biggl(-\frac{\beta-\tau}{2m\omega^2}z^2\biggr)
E_{\nn}\biggl\{V(y(\tau)-y(0))\times{}
$$
$$
\times\exp\biggl(-\int_0^\tau V(y(s)-y(0))\,ds+
iz(y(\tau)-y(0))\biggr)\biggr\}=
$$
$$
=\exp\biggl(-\frac{\beta-\tau}{2m\omega^2}z^2\biggr)
\int_{-\infty}^\infty d\eta\, V(\eta)e^{iz\eta}
E_{\nn}\biggl\{\exp\biggl(-\int_0^\tau V(y(s)-y(0))\,ds\biggr)
\delta_{\tau,\eta}(x)\biggr\},
$$
which follows from Theorem~6 and the properties of function~(63).
Therefore,
$$
u(\beta,\xi)=E_{\nn}\{\delta_{\beta,\xi}(x)\}-
\frac1{2\pi}\int_0^\beta d\tau\int_{-\infty}^\infty dz\,e^{-iz\xi}
\exp\biggl(-\frac{\beta-\tau}{2m\omega^2}z^2\biggr)\times
$$
$$
\times\int_{-\infty}^\infty d\eta \, V(\eta)e^{iz\eta}u(\tau,\eta).
$$
In view of
$$
\int_{-\infty}^\infty dz\,\exp\biggl(-iz\xi-\frac{\beta-\tau}{2m\omega^2}
z^2+iz\eta\biggr)=\sqrt{\frac{2\pi m\omega^2}{\beta-\tau}}
\exp\biggl[-\frac{2m\omega^2}{\beta-\tau}\frac{(\xi-\eta)^2}4\biggr],
$$
we use the lemma to obtain
$$
u(\beta,\xi)=\sqrt{\frac{m\omega^2}{2\pi\beta}}
\exp\biggl(-\frac{m\omega^2}{2\beta}\xi^2\biggr)-{}
$$
$$
-\int_0^\beta\int_{-\infty}^\infty V(\eta)u(\tau,\eta)
\sqrt{\frac{m\omega^2}{2\pi(\beta-\tau)}}
\exp\biggl[-\frac{m\omega^2}{2(\beta-\tau)}(\xi-\eta)^2\biggr]\,d\eta\,d\tau.
\eqno(68)
$$
Direct verification now readily shows that function~(68) satisfies
Eq.~(67). The corresponding initial and boundary conditions are
obviously satisfied. Theorem~7 is proved.

\section { Inequalities for equilibrium averages}

We consider a system with a Hamiltonian
$$
\widehat H=\widehat\Gamma+\widehat V,
$$
where $\widehat V=V(\hat q)$ is an interaction term, and also a
one-dimensional family of Hamiltonians,
$$
\widehat H(h)=\widehat\Gamma(h)+\widehat V,\quad h\in R,
$$
$$
\widehat\Gamma(h)=\frac{\hat p^2}{2m}+\frac{m\omega^2}2(\hat q-h)^2.
$$
The statistical sum
$$
Z(h)=\tr e^{-\beta\widehat H(h)}
$$
of the system under consideration becomes
$$
Z(h)=\tr e^{-\beta[\widehat\Gamma+V(\hat q+h)]}
$$
after the canonical transformation $\hat q-h\to\hat q$. We
assume that the interaction potential is nonnegative and symmetric, i.e.,
$$
V(x)\geq0,\qquad V(x)=V(-x).
$$
Using the chronological-ordering operator, we can write~\cite{15a2}
$$
e^{-\beta(\widehat{\Gamma}+\widehat{V})}=e^{-\beta\widehat{\Gamma}}T
\exp\biggl(-\int_0^{\beta}ds\,e^{s\widehat{\Gamma}}
\widehat Ve^{-s\widehat{\Gamma}}\biggr).
$$
Therefore,
$$
R(h)\equiv\frac{\tr e^{-\beta
\widehat{H}(h)}}{\tr e^{-\beta\widehat{\Gamma}}}=
\biggl\langle T\exp\biggl[-\int_0^{\beta} ds\,V
\bigl(\widehat Q(s)+h\bigr)\biggr]\biggr\rangle_{\widehat{\Gamma}}.
\eqno(69)
$$
Expressing relation~(69) via the Bogolyubov functional
integral, we obtain
$$
R(h)=\int_X\exp\biggl[-\int_0^{\beta}ds\,
V\bigl(x(s)+h\bigr)\biggr]\,d\nn(x).
$$

We now apply the theorem on a linear change of variable in an integral with
respect to a Gaussian measure~\cite{11a3}. This gives
$$
\int_XF(x)\,d\mu(x)=e^{-\|a\|_H^2/2}\int_XF(x+a)e^{-(a,x)}\,d\mu(x)
\eqno(70)
$$
for an integrable functional $F(x)$ and a function $a\in H$. In this
situation, we use formula~(70) for the case of the Bogolyubov measure
and the constant functions $a$ that belong to $H$. This results in
$$
\int_XF(x+a)\,d\nn(x)=e^{-\beta m\omega^2a^2/2}\int_XF(x)
\exp\biggl\{am\omega^2\int_0^{\beta}x(t)\,dt\biggr\}\,d\nn(x).
$$
This relation permits writing the function $R(h)$ in the form
$$
R(h)=e^{-\beta m\omega^2h^2/2}\int_X
\exp\biggl\{-\int_0^{\beta}V\bigl(x(t)\bigr)\,dt\biggr\}
\exp\biggl\{mh\omega^2\int_0^{\beta}x(t)\,dt\biggr\}\,d\nn(x).
$$

We now consider the Fourier--Gauss transform
$$
\tilde f(y)\equiv F(f;y)=\int_Xf(x+iy)\,d\nn(x)
$$
of a functional $f(x)$ and the Parseval relation
$$
\int_Xf\left(\frac x{\sqrt2}\right)g^*
\left(\frac x{\sqrt2}\right)\,d\nn(x)=
\int_XF\left(f;\frac y{\sqrt2}\right)
F^*\left(g;\frac y{\sqrt2}\right)\,d\nn(y)
\eqno(71)
$$
for the the case of functionals
$$
f(x)=F(x)\equiv\exp\biggl\{-\int_0^{\beta} dt\,V\bigl(x(t)\bigr)\biggr\}
$$
and
$$
g(x)=\exp\biggl\{mh\omega^2\int_0^{\beta}x(t)\,dt\biggr\}.
$$
Relation~(71) becomes
$$
e^{-\beta mh^2\omega^2/2}\int_XF\left(\frac x{\sqrt2}\right)
\exp\biggl\{\frac1{\sqrt2}hm\omega^2\int_0^{\beta}x(t)\,dt\biggr\}\,
d\nn(x)=
$$
$$
=\int_X\widetilde F\left(\frac y{\sqrt2}\right)\exp\biggl\{
\frac i{\sqrt2}hm\omega^2\int_0^{\beta}y(t)\,dt\biggr\}\,d\nn(y),\qquad
$$
whence we see that if the inequality
$$
\widetilde F(y)\geq0
\eqno(72)
$$
holds for all $y$, then
$$
R(h)=\widetilde F(-ih)\leq R(0)=\widetilde F(0).
\eqno(73)
$$
Condition~(72) is proved as follows. We have
$$
\widetilde F(y)=e^{(y,y)/2}\int_X
\exp\biggl\{-\int_0^{\beta} V(x)\,dt+i(x,y)\biggr\}\,d\nn(x)=
$$
$$
=e^{(y,y)/2}\int_X
\exp\biggl\{-\int_0^{\beta}V(x)\,dt-i(x,y)\biggr\}\,d\nn(x)
=\widetilde F^*(y)
$$
for symmetric functionals, i.e., the Fourier--Gauss transform is real in this
case.
We prove that it is nonnegative.
In view of
$$
e^{\|y\|^2/2}\int_Xe^{i(x,y)}\,d\nn(x)=1,
$$
applying the Jensen inequality yields
$$
\widetilde F(y)\geq\exp\biggl\{-e^{\|y\|^2/2}\int_Xd\nn(x)\,
e^{i(x,y)}\int_0^{\beta} V\bigl(x(t)\bigr)\,dt\biggr\}\geq0,
$$
which precisely completes the proof of~(72).

In particular, condition~(73) implies that
$$
(\hat q,\hat q)_{\widehat{H}}\leq\frac1{\beta m\omega^2},
\eqno(74)
$$
where the Bogolyubov inner product of arbitrary operators
$\widehat A$ and $\widehat B$ is defined as
$$
(\widehat A,\widehat B)_{\widehat{H}}
=\frac1{\beta\tr e^{-\beta\widehat{H}}}
\int_0^{\beta} ds\,\tr\bigl
[e^{-s\widehat{H}}\widehat Ae^{-(\beta-s)\widehat{H}}
\widehat B\bigr]=(\widehat B,\widehat A)_{\widehat{H}}.
$$

Using the relations
$$
\hat q=\frac1{\sqrt{2m\omega}}(\hat b+\hat b^{\dag}),\qquad
\hat p=i\sqrt{\frac{m\omega}2}\,\bigl(\hat b^{\dag}-\hat b\bigr)
$$
to pass from the operators $\hat q$ and $\hat p$ to $\hat b$ and
$\hat b^{\dag}$ and taking the selection rules for equilibrium
averages with respect to the quadratic Hamiltonian into account, we bring
inequality~(74) to the form
$$
\bigl(\hat b^{\dag},\hat b\bigr)_{\widehat{H}}\leq{(\beta\omega)}^{-1}.
$$

Relation~(74) can be used to derive an inequality for the Gibbs
equilibrium average $\langle{\hat q}^2\rangle_{\widehat{H}}$.
For this, the Falk--Bruch inequality~\cite{26a2} should be used. Let
$$
g=\langle{\hat q}^2\rangle_{\widehat{H}},\qquad b
=(\hat q,\hat q)_{\widehat{H}}, \qquad
c=\bigl\langle\bigl[\hat q, [\beta\widehat H,\hat
q]\bigr]\bigr\rangle_{\widehat{H}}.
$$
We also assume that the upper
estimates $b\leq b_0$ and $c\leq c_0$ hold.  Then
$$
g\leq g_0\equiv\frac12\sqrt{c_0b_0}\,\coth\sqrt{\frac{c_0}{4b_0}}.
$$
In the case under study, we have $b_0=(\beta m\omega^2)^{-1}$ and
$c_0=\beta/m$, and the above inequality gives
$$
\langle{\hat q}^2\rangle_{\widehat{H}}\leq
\frac1{2m\omega}\coth\frac{\beta\omega}2= \langle{\hat
q}^2\rangle_{\widehat{\Gamma}}.
\eqno(75)
$$

Condition~(73) is an example of the so-called Gaussian domination
condition~\cite{27a2}, and condition~(74) implied
by~(73) is an example of the so-called local Gaussian domination
condition~\cite{28a2}, which plays an important role in phase transition
theory. An estimate of type~(75) was previously found for a less
general case of a one-dimensional nonlinear oscillator~\cite{29a2}.

Partial financial support by
RFBR, grant 99--01--00887 (Russia).

\end{document}